\documentclass[iop,twocolappendix,appendixfloats]{emulateapj}
\usepackage{apjfonts,amsmath}
\usepackage{color}

\newcommand\beq{\begin{equation}}
\newcommand\eeq{\end{equation}}

\newcommand{\ca}{{\sc camira}\ }

\shorttitle{Cluster Galaxy Evolution with HSC}
\shortauthors{Lin et al.}

\begin{document}


\title{First results on the cluster galaxy population from the Subaru Hyper Suprime-Cam survey.  III.~Brightest cluster galaxies, stellar mass distribution, and active galaxies}

\author{
Yen-Ting Lin\altaffilmark{1},
Bau-Ching Hsieh\altaffilmark{1},
Sheng-Chieh Lin\altaffilmark{1,2},
Masamune Oguri\altaffilmark{3,4},
Kai-Feng Chen\altaffilmark{1,5},
Masayuki Tanaka\altaffilmark{6},\\
I-Non Chiu\altaffilmark{1},
Song Huang\altaffilmark{7,4},
Tadayuki Kodama\altaffilmark{8},
Alexie Leauthaud\altaffilmark{7,4},
Surhud More\altaffilmark{4},
Atsushi J.~Nishizawa\altaffilmark{9},
Kevin Bundy\altaffilmark{7,4},
Lihwai Lin\altaffilmark{1},
and Satoshi Miyazaki\altaffilmark{6}
}

\altaffiltext{1}{Institute of Astronomy and Astrophysics, Academia Sinica, Taipei 10617, Taiwan; ytl@asiaa.sinica.edu.tw}
\altaffiltext{2}{Institute of Astrophysics, National Taiwan University}
\altaffiltext{3}{Department of Physics and Research Center for the Early Universe, The University of Tokyo}
\altaffiltext{4}{Kavli Institute for the Physics and Mathematics of the Universe, The University of Tokyo}
\altaffiltext{5}{Department of Physics, National Taiwan University}
\altaffiltext{6}{National Astronomical Observatory of Japan}
\altaffiltext{7}{Department of Astronomy and Astrophysics, University of California, Santa Cruz}
\altaffiltext{8}{Astronomical Institute, Tohoku University}
\altaffiltext{9}{Institute for Advanced Research, Nagoya University}

\begin{abstract}

The unprecedented depth and area surveyed by the Subaru Strategic Program with the Hyper Suprime-Cam (HSC-SSP) have enabled us to construct and publish the largest distant cluster sample out to $z\sim 1$ to date.  In this exploratory study of cluster galaxy evolution from $z=1$ to $z=0.3$, we investigate the stellar mass assembly history of brightest cluster galaxies (BCGs),  evolution of stellar mass and luminosity distributions, stellar mass surface density profile, as well as the population of radio galaxies.  Our analysis is the first high redshift application of the {\it top N} richest cluster selection, which is shown to allow us to trace the cluster galaxy evolution faithfully.  
Over the 230\,deg$^2$ area of the current HSC-SSP footprint, selecting the {\it top 100}  clusters in each of the 4 redshift bins allows us to observe the buildup of  galaxy population in descendants of clusters whose $z\approx 1$ mass is about $2\times 10^{14}\,M_\odot$.
Our stellar mass is derived from a machine-learning algorithm, which is found to be unbiased and accurate with respect to the   COSMOS data.
We find very mild stellar mass growth in BCGs (about 35\% between $z=1$ and 0.3), and no evidence for evolution in both the total stellar mass--cluster mass correlation and the  shape of the stellar mass surface density profile.
We also present the first measurement of the radio luminosity distribution in clusters out to $z\sim 1$, and show hints of changes in the dominant accretion mode powering the cluster radio galaxies at $z\sim 0.8$.

\end{abstract}


%%%%%%%%%%%%%%%%%%%%%%%%%%%%%%%%%%%%%%%%%%%%%
%%%%%%%%%%%%%%%%%%%%%%%%%%%%%%%%%%%%%%%%%%%%%
\section{Introduction} 
\label{sec:intro}

The stark difference in galaxy populations between galaxy clusters and the ``field'' has long been recognized \citep[e.g.,][]{dressler80}.  It is important to understand how  quiescent early type galaxies come to dominate the  galaxy population as seen in  present-day clusters.  One approach is to compare the galaxy populations in galactic systems of different halo masses (e.g., field, groups, and clusters), so that the relative importance of processes that depend on the total mass of the systems could be estimated \citep[e.g.,][]{treu03,lin04,tanaka05,koyama07,vanderburg14}. 
Ideally, for such comparisons, one needs to take into account the fact that structures grow hierarchically (e.g., progenitors of a present-day massive cluster are lower mass clusters at higher redshift) and consider galactic systems across a wide range in cosmic time.
Another argument for folding in the time dimension in such studies is that
cluster galaxies are believed to have experienced accelerated evolution with respect to the field population; at some point in the past, when the statistical properties of cluster galaxies are closer to that in the field, we may then be able to identify environmental factors that are involved in shaping the cluster galaxy population \citep[e.g.,][]{hayashi10,tran10,brodwin13}.

Our current knowledge of cluster galaxy evolution has been largely built upon the combination of massive, local cluster samples from the 
ROSAT All-Sky Survey (e.g., \citealt{ebeling98,boehringer00,boehringer04b}), 
Sloan Digital Sky Survey (SDSS; \citealt{york00}) and high-$z$ cluster surveys carried out over areas up to tens of square degrees \citep[e.g.,][]{brodwin13,vanderburg15}.
With the advent of cluster surveys via the Sunyaev-Zel'dovich effect (SZE; \citealt{sunyaev70}) such as the Atacama Cosmology Telescope (ACT; \citealt{sievers13}), South Pole Telescope (SPT; \citealt{carlstrom11}), and Planck \citep[][]{planck13xxix}, distant cluster samples detected over hundreds of square degrees or larger area have finally become available.  These are valuable in studying galaxies in ``mature'' or extreme galactic systems as these SZE surveys generally detect the most massive clusters  \citep[][]{hilton13,zenteno16,chiu16b}.
Only until very recently, cluster searches with deep, wide optical surveys such as the Dark Energy Survey and Subaru HSC-SSP Survey are beginning to provide cluster samples that cover wide ranges in both redshift and mass, thus enabling comprehensive studies of cluster galaxy evolution for the first time \citep[e.g.,][]{rykoff16,hennig17}.

As alluded above, in order to study the evolution of galaxies in clusters in the context of hierarchical structure buildup, one needs to construct cluster samples that could reasonably be regarded as representing a progenitor-descendant relationship.  
A few studies have been carried out following such an idea, mainly devoted to understanding the way BCGs acquire their stellar mass over cosmic time \citep[e.g.,][]{lidman12,lin13,inagaki15,zhang16}.

In this paper we present an exploratory analysis of the evolution of the cluster galaxy population from $z\approx 1$ to $z=0.3$, using the initial cluster sample from the HSC-SSP survey, which is constructed by utilizing the multi-color red-sequence algorithm \ca  \citep{oguri17}.  In each of the 4 redshift bins that occupy equal comoving volume, we focus on the top 100 richest clusters.  
Such a {\it ``top N''} cluster selection provides an efficient and reproducible way to construct cluster samples that can be regarded as statistically representing a progenitor-descendant relationship, as has been demonstrated and used in the literature \citep[e.g.,][]{inagaki15}.
We emphasize that this progenitor-descendant relationship is strictly statistical in nature, and is only meant to suggest that the ensemble properties of the higher redshift sample should be similar to that of the progenitors of the lower redshift sample.  
In practice, we make use of the richness as the mass proxy to select the top $N$ most massive clusters at a given redshift. 
In principle, with a low intrinsic scatter of the richness--cluster mass relation and a large survey volume, both of which could be realized after the HSC-SSP survey enters the mature phase, the {\it top N} selection should
provide a statistical perspective for comparison between  progenitors and descendants.

In addition to studying the stellar mass assembly history of BCGs, we  stack photometric data from the clusters and examine the evolution of the stellar mass distribution (SMD), the $i$-band luminosity distribution (LD), the surface stellar mass density profiles, as well as the fraction of galaxies that are active in radio, all using a statistical background subtraction method.

This is the third paper in a series where we have studied the evolution of galaxy population in clusters detected in the HSC-SSP survey.  In \citet{jian17} the environmental dependence of quenching mechanisms of galaxies is studied.  In \citet{nishizawa17} we have measured the radial density profiles of red and blue galaxies, as well as the intrinsic scatter in color of the red sequence, finding that the scatter is almost constant down to $z=24$ mag.

The structure of this paper is as follows.  An overview of our analysis is given in Section~\ref{sec:overview}, where we describe our cluster and galaxy samples,  demonstrate the validity of both our background subtraction method and the {\em top N} cluster selection scheme, and estimate the typical mass of our cluster sample.  Our results are presented in Section~\ref{sec:res}.  
After discussing potential ways to improve the techniques used in the current analysis in Section~\ref{sec:disc}, we summarize our results in Section~\ref{sec:summary}.

Throughout this paper we adopt a {\it WMAP5} \citep{komatsu09} $\Lambda$CDM cosmological model,
where $\Omega_m=0.26$, $\Omega_\Lambda=0.74$, $H_0=100h~{\rm km\,s^{-1}\,Mpc^{-1}}$ with $h=0.71$. 
Unless otherwise noted, the halo mass definition we adopt is $M_{200c}$, the mass enclosed in $r_{200c}$, within which the mean overdensity is 200 times the critical density of the universe at the redshift of the halo.  For simplicity, we omit the letter $c$ in mass, radius, and concentration.  Where needed, the mass and radius $M_{500}$ and $r_{500}$ are defined in a similar fashion.
All magnitudes are in the AB system.

%%%%%%%%%%%%%%%%%%%%%%%%%%%%%%%%%%%%%%%%%%%%%
%%%%%%%%%%%%%%%%%%%%%%%%%%%%%%%%%%%%%%%%%%%%%
\section{Analysis Overview}
\label{sec:overview}

In this section we first describe the HSC-SSP survey, then the cluster and galaxy samples from the  Survey used for our analysis, and provide details of the estimation of  stellar mass, the stacking method, and the background subtraction scheme used to construct the SMD and LD in clusters.  We verify these methods with mock galaxy catalogs, as described in Section~\ref{sec:mice}.  In Section~\ref{sec:topn} we use an $N$-body simulation to justify the {\it top N} cluster selection approach in tracing the cluster evolution, while in Section~\ref{sec:mass} we use two different ways to estimate  the mass of our cluster sample.

%%%%%%%%%%%%%%%%%%%%%%%%%%%%%%%%%%%%%%%%%%%%%
\subsection{The HSC-SSP Survey}

The HSC-SSP survey \citep{aihara17b} is one of the Subaru Strategic Programs, which are designed to enable large scale projects to be conducted with the facility instruments of the Subaru Telescope.  
The 300-night survey is carried out
using the wide-field Hyper Suprime-Cam \citep{miyazaki17b}
 by a collaboration of astronomers from Japan, Taiwan, and Princeton University, and consists of a wide, a deep, and an ultradeep layer.
Each layer is observed in the $grizy$ broad-band filters.  There are also narrow-band observations in the deep and ultradeep layers. 
The wide layer reaches to a depth of $r\sim 26$ mag over $1400\,$deg$^2$.
In the deep and ultradeep layers, the target depth and area are $r\sim 27$ mag, $27\,$deg$^2$ (over 4 separate fields) and $r\sim 28$ mag, $3.5\,$deg$^2$ (two fields: Subaru-XMM Deep Field and COSMOS), respectively.  
The primary science goals of the survey are to constrain the properties of dark matter and dark energy via the cosmic structure growth and expansion history derived from weak lensing tomography and type Ia supernovae, and to trace galaxy and active galactic nuclei (AGN) evolution from the local Universe all the way to the epoch of reionization.

The survey data is reduced by a pipeline (\citealt{bosch17}) derived from that developed for the Large Synoptic Survey Telescope (\citealt{ivezic08}).  Astrometric and photometric calibrations are carried out by comparison with data from the PanSTARRS1 survey (\citealt{chambers16}).
The first public  release of the HSC-SSP survey data is presented in \citet{aihara17}.

%%%%%%%%%%%%%%%%%%%%%%%%%%%%%%%%%%%%%%%%%%%%%
\subsection{Cluster and Galaxy Samples}
\label{sec:sample}

The \ca algorithm \citep{oguri14} is run on the ``S16A'' internal data release of the HSC-SSP survey \citep{aihara17}, covering roughly $230\,$deg$^2$ observed in all five filters.  The resulting 1921 clusters above the richness limit $\hat{N}>15$ span the redshift range of $z=0.1-1.1$ \citep{oguri17}.  
The richness is defined to be the number of red member galaxies with stellar mass $M_{\rm star}\ge 10^{10.2}\,M_\odot$ lying within a physical radius of $\approx 1.4\,$Mpc.
Based on the abundance of clusters, the richness limit $\hat{N}=15$ roughly corresponds to $M_{200}\approx 1.3\times 10^{14}\,M_\odot$.
Comparisons with spectroscopic catalogs and existing X-ray clusters indicate that the photometric redshifts $z_{ph,c}$ of the clusters are quite accurate (with bias and scatter of $-0.0013$ and $0.0081$, respectively) and the richness is a good proxy of cluster mass.
Although \ca produces the member catalogs for each of the clusters, they only contain red member galaxies.  Since we would like to understand the evolution of blue populations as well, we do not make use of these member catalogs directly; rather, we adopt a statistical background subtraction approach that can be applied to both populations. 

The photometric catalog we use to measure the cluster galaxy properties is the same one used in the cluster detection.  The catalog is limited to {\tt cmodel} magnitude $z_{\rm cmodel}<24$.  Additional magnitude limits of $r_{\rm cmodel}<26.5$ and $i_{\rm cmodel}<26$, as well as various flags, are applied to ensure clean photometry (please refer to \citealt{oguri17} for more details).  
All magnitudes are corrected for Galactic extinction.
Star-galaxy separation is performed using  $i$-band measurements as this band is of the best imaging quality.  Unless noted explicitly, {\tt cmodel} $grizy$ measurements are used throughout our analysis.

As described in \citet{aihara17} and \citet{bosch17}, the version of the HSC pipeline  used to produce the S16A release occasionally generates problematic photometry in crowded fields such as cluster centers.  As the pipeline has a hard time deblending galaxies in such regions, magnitudes of the deblended objects could be estimated erroneously.  As a remedy, galaxy colors are measured on the objects in the un-deblended images after the point spread function (PSF) sizes are matched for all five broad-bands; the PSF sizes are degraded to $1.1''$, and the aperture size of $1.1''$ in diameter is used for the color measurement.  We still use the {\tt cmodel} magnitudes in the $z$-band for the flux measurement.

\begin{figure}
\epsscale{1.5}
\plotone{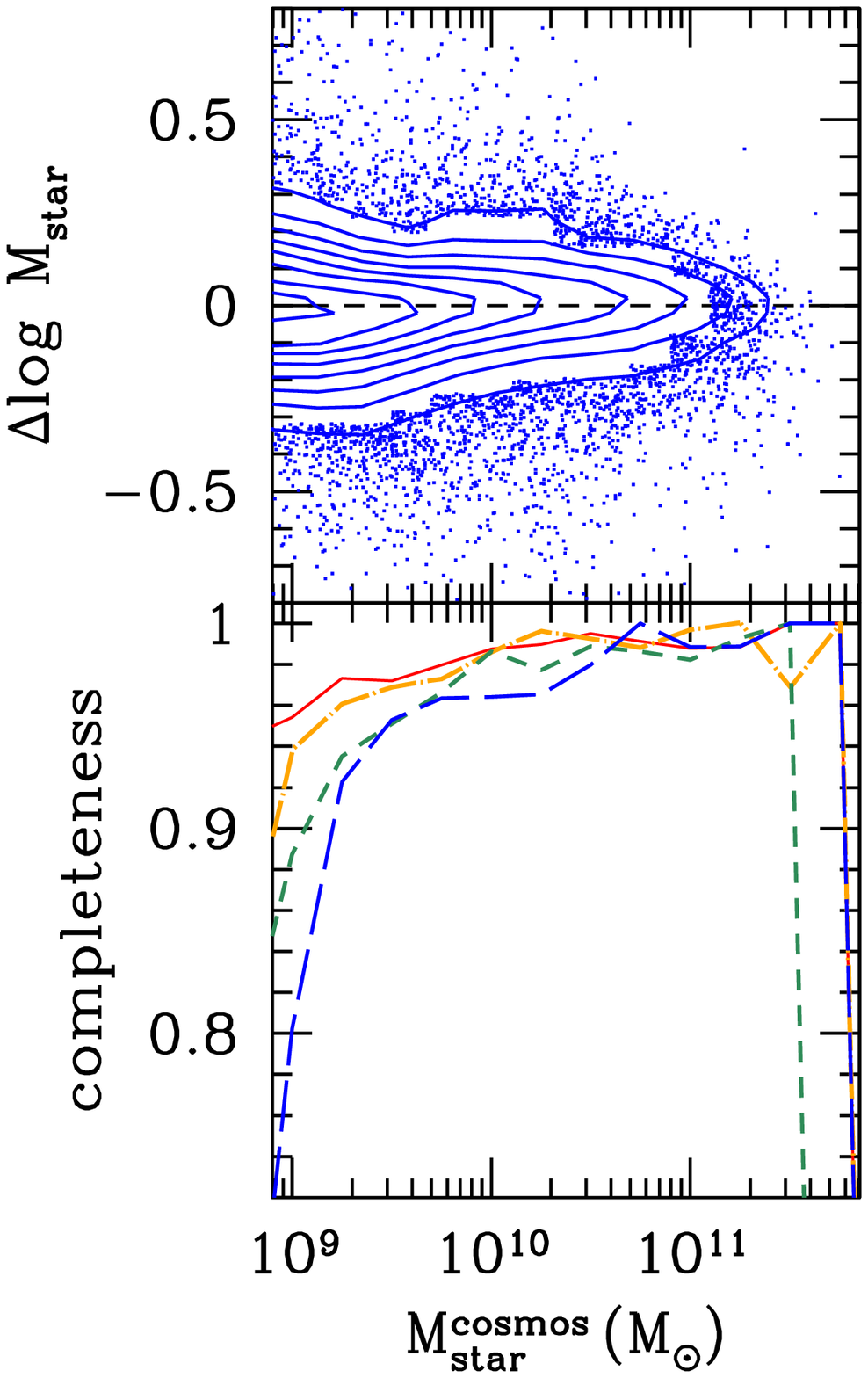}
\vspace{-5mm}
\caption{ 
The top panel shows the difference in logarithm between the {\sc demp}-derived stellar mass and the stellar mass in the COSMOS2015 catalog \citep{laigle16},
$\Delta \log M_{\rm star} \equiv \log M^{\rm hsc}_{\rm star}-M^{\rm cosmos}_{\rm star}$, as a function of COSMOS2015 stellar mass.  
Our stellar mass is unbiased with respect to the COSMOS2015 results.
This comparison is done for galaxies with COSMOS2015 photometric redshift $z_{ph}=0.7-1$, although for galaxies at lower redshifts the performance is similar.  The lower panel shows the completeness in 4 redshift bins (blue: $z_{ph}=0.9-1.02$; green: $z_{ph}=0.77-0.9$; orange: $z_{ph}=0.6-0.77$; red: $z_{ph}=0.3-0.6$).  We achieve very high completeness for galaxies more massive than a few times $10^9\,M_\odot$.
}
\label{fig:comp}
\end{figure}

The stellar mass and luminosity of  galaxies are estimated via the machine learning algorithm {\sc demp} \citep[Direct Empirical Photometric method;][]{hsieh14}. 
The training set is produced by cross matching the HSC {\it ultradeep} layer data in the COSMOS field with the COSMOS2015 catalog \citep{laigle16}, which we take as the ``truth'' table,
as the stellar mass and photometric redshift therein are derived robustly using 30 photometric bands ranging from near-UV to 24\,$\mu m$, including deep data from UltraVISTA \citep{muzzin13} and SPLASH \citep{steinhardt14}.
The exquisite data quality of the HSC ultradeep layer (reaching to $5\sigma$ depth of $i\sim 27.2$ for point sources; \citealt{aihara17}) ensures an accurate mapping between the observed $grizy$ magnitudes and the stellar mass (or luminosity) as a function of redshift.
To validate our training, we compare the {\sc demp}-derived stellar mass ($M^{\rm hsc}_{\rm star}$), based on the {\it wide-layer-depth} data in the COSMOS field\footnote{As is shown in \citet{tanaka17}, the photometric data in the ultradeep layer and the wide-layer-depth stack in the COSMOS field have negligible correlation and thus can be regarded as independent datasets.  We have also used 90\% of the ultradeep data for training, and wide-layer-depth objects that are matched to the other 10\% in the ultradeep for validation, and found essentially the same results.}, with that from the \citet{laigle16} catalog ($M^{\rm cosmos}_{\rm star}$).  The result for galaxies at $z=0.7-1$ is shown in the top panel of Figure~\ref{fig:comp}.
It is clearly shown that our stellar mass is unbiased with respect to the COSMOS masses (mean of $\Delta \log M_{\rm star} \equiv \log M^{\rm hsc}_{\rm star}-M^{\rm cosmos}_{\rm star}$ is $-0.02$), with a scatter of 0.2\,dex.  The comparisons done for lower redshift galaxies show similar results.
We note in passing that for galaxies at $z>0.8$ or so, the $grizy$ photometry from the HSC 
does not sample much of the galaxy spectral energy distribution (SED) in the restframe optical,
and therefore SED-fitting based methods using solely  the HSC photometry are prong to larger uncertainties and/or biases for high-$z$ galaxies compared to the empirical approach adopted here.

In the lower panel of Figure~\ref{fig:comp}, we examine the completeness of our stellar mass estimates.  
The differential completeness is defined for galaxies within a given mass range ($\log M-\Delta/2$ to $\log M+\Delta/2$) to be 
\begin{equation}
C(\tilde{M}) = \frac{ N(\tilde{M}^{\rm cosmos}_{\rm star} \in \{ \tilde{M}-\Delta/2, \tilde{M}+\Delta/2 \},\ \tilde{M}^{\rm hsc}_{\rm star}\in \{6,14 \}) } {N(\tilde{M}^{\rm cosmos}_{\rm star} \in \{ \tilde{M}-\Delta/2, \tilde{M}+\Delta/2 \}) } ,
\end{equation}
where a tilde denotes a quantity in logarithm, 
the denominator is the number of all galaxies detected in the COSMOS2015 catalog with $M^{\rm cosmos}_{\rm star}$ in that mass range, while the numerator is the same, except we also require the {\sc demp}-derived stellar mass to be in the range $10^6-10^{14}\,M_\odot$ (a condition that removes about $6\%$ of the galaxies;  restricting the {\sc demp}-derived stellar mass to be within 2\,dex of the COSMOS2015 value results in essentially the same completeness curves).
In practice, we choose $\Delta=0.25\,$dex in generating the Figure.
The 4 curves in the panel show the completeness for galaxies in 4 redshift bins used in our analysis.  Generally speaking, our completeness is $>95\%$ at all redshifts for $M^{\rm cosmos}_{\rm star}\ge 3\times 10^9\,M_\odot$.

Now that we have shown that {\sc demp} could recover stellar masses in a unbiased and accurate fashion using the HSC wide layer photometry, 
for  galaxies likely associated with a cluster (see Section~\ref{sec:mice} below), 
we derive the stellar mass using {\sc demp}. 
We note that, as \citet{laigle16} adopt the \citet{chabrier03} initial mass function (IMF), our {\sc demp}-based masses inherit the same assumption on the IMF.
Finally, following essentially the same procedure, {\sc demp} is trained with the COSMOS2015 catalog as well as the HSC ultradeep data to derive the (rest-frame) $i$-band absolute magnitude.

A part of our analysis of cluster galaxy evolution concerns with the radio luminosity distribution (RLD) and the fraction of galaxies that are active in the radio wavelength, where the radio-active galaxies are simply defined to be above certain threshold in radio luminosity.  The identification of radio-active galaxies is carried out by matching our galaxy catalog with the source catalog from the FIRST (Faint Images of the Radio Sky at Twenty-Centimeters; \citealt{becker95}) survey, which has a $5\sigma$ flux limit of 1\,mJy.  We note the HSC-SSP survey area is entirely within the FIRST footprint.
We consider a match if an HSC galaxy has a counterpart in the FIRST catalog within $1''$.  For BCGs, we have visually inspected the matching results, and take proper flux measurements when the radio counterpart is an obvious multi-component source.

%%%%%%%%%%%%%%%%%%%%%%%%%%%%%%%%%%%%%%%%%%%%%
\subsection{Composite Stellar Mass Distributions and Surface Density Profiles}
\label{sec:mice}

To construct the composite SMD, we proceed as follows.   
For a given cluster, we assume all galaxies lying within a projected physical distance of $r_{\rm max}$ to be at the cluster redshift, and estimate the stellar mass using {\sc demp}.  
The ``apparent'' stellar mass distribution is obtained by simply counting number of galaxies as a function of stellar mass, for all galaxies within a projected distance $r_{\rm cl}$.  This obviously has contributions from both cluster members and foreground/background galaxies.  The latter is estimated by an annulus with inner and outer radii $r_{\rm an,in}$ and $r_{\rm max}$.  For all the clusters in a given redshift bin, we sum over both the apparent stellar mass distribution and the distribution in the background annulus and then subtract the latter from the former (accounting for differences in the area).  
After further normalized by the number of clusters in the redshift bin, we obtain the SMD.

In our analysis, we adopt $r_{\rm an,in}=5\,$Mpc and $r_{\rm max}=7$\,Mpc, and $r_{\rm cl}=r_{200}$, our best estimate of the virial radius of the cluster (see Section~\ref{sec:topn}). 
It is found that as long as $r_{\rm an,in}$ is sufficiently large, our results do not sensitively depend on the exact choice of the values (e.g., \citealt{lin07}).
Some clusters are close to the boundary of the survey and thus the annulus region is not completely covered by the survey data; others have big holes in the annulus region due to bright stars or other data reduction issues.  These clusters are excluded when we construct the composite SMDs (see more discussion in Section~\ref{sec:smf}).

To test if the above procedure can unbiasedly uncover the true SMD, we use a set of mock cluster and galaxy catalogs, which are based on the MICE mock catalog (\citealt{carretero15}).
The MICE mock is produced by populating halos in a lightcone simulation with galaxies using a combination of halo occupation distribution and abundance matching techniques.  The algorithm is tuned to reproduce the observed luminosity function, color-magnitude relation, and galaxy clustering properties in the local universe.
The mock catalog provides apparent magnitudes in Dark Energy Survey (DES) $grizY$ and VISTA $JHK$ filters for galaxies out to $z=1.4$ over 1 octant of the sky, and is complete to $r$-band absolute magnitude $M_r\le -18.9$.

For our purpose, we have extracted catalogs of galaxies projected within $r_{\rm max}$ around the top 100 most massive halos at $z=0.9-1.02$ over a $200\,$deg$^2$ area from the full MICE mock, and have treated the mock galaxies the same way as we do on real galaxies, that is, assuming they all lie at the redshift of their respective halos, and estimating the stellar mass with the SED fitting technique, using the software package {\sc NewHyperz}\footnote{\url{http://userpages.irap.omp.eu/~rpello/newhyperz/}} with the \citet[][hereafter BC03]{bruzual03} templates\footnote{Although the stellar mass is derived in a different fashion from our real data, the point of this exercise is to see whether the statistical background subtraction method allows us to recover the true SMD.  The methodology for estimating stellar masses is not relevant in this context.}.
We also apply survey masks and artificially create boundary effects to mimic observational defects at the catalog level.

\begin{figure}
\epsscale{0.9}
\plotone{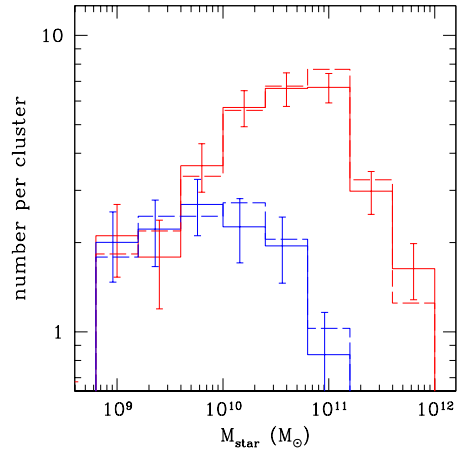}
\vspace{-2mm}
\caption{ 
The underlying and recovered SMDs, shown as dashed and  solid histograms, respectively, based on tests with the MICE mock catalog.  The red and blue histograms show the results for red and blue galaxies separately.
The agreement suggests that our background subtraction scheme works well in providing an unbiased estimate of the true SMD.
}
\label{fig:mice}
\end{figure}

Figure~\ref{fig:mice} shows the performance of our method in recovering the SMD.  While the dashed histogram shows the true SMD in the mocks
(constructed from galaxies lying within the projected virial radius, and with redshifts within $\Delta z=0.005$ from the halos, chosen to approximate what is attainable with spectroscopic redshifts\footnote{Even with spectroscopic redshifts, one cannot completely remove projections from the correlated structures around the clusters along the line-of-sight.  Comparing the number of  mock galaxies truly associated with the halos with that from the ``ideal'' case attainable from spectroscpy (i.e., with $\Delta z\le 0.005$), we find that the surrounding structures contribute about 20\% of the galaxies in projection, which is consistent with the findings from a rigorous estimation for redMaPPer clusters (T.~Sunayama et al.~in prep.).}), 
the solid one is obtained with the background subtraction scheme.  We see that over the mass range of interest (e.g., $>10^9 M_\odot$), our method works fairly well.
We have repeated this exercise in another redshift range ($z=0.6-0.77$), also finding 
excellent agreement between 
the underlying and recovered SMDs.

\begin{figure}
\epsscale{1.4}
\plotone{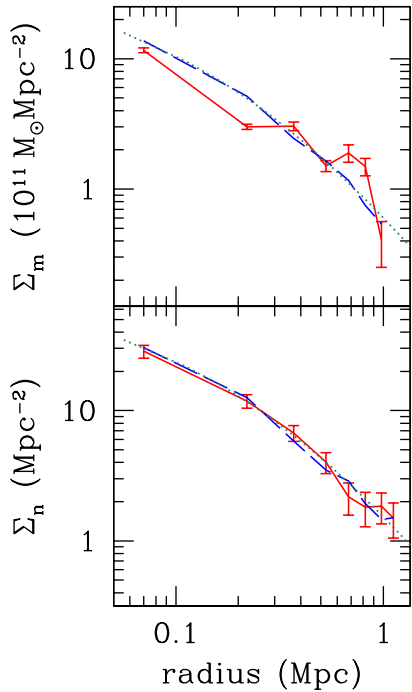}
\vspace{-9mm}
\caption{ 
{\it Top:} a comparison of the true and recovered stellar mass surface density profiles, again using the MICE mock catalog.  The blue dashed curve shows the true profile, while the solid curve with error bars represent the recovered profile.  The dotted curve shows an NFW profile with a concentration $c=4$, scaled to metric radius using the median $r_{200}=1.12$\,Mpc of the top 100 halos. {\it Bottom:} similar to the top panel, but for the surface number density profile (the NFW fit now is with $c=3.5$).  It appears that the surface number density profile can be better recovered than the stellar mass density profile.
}
\label{fig:mice2}
\end{figure}

We would also like to examine the spatial distribution of  stellar mass in clusters, again by stacking clusters and employing the statistical background subtraction scheme.  This is  tested with the MICE mocks as well.  Figure~\ref{fig:mice2} (top panel) shows, as the dashed curve, the true stellar mass surface density profile, and the solid curve is the derived profile, for galaxies more massive than $10^{10}\,M_\odot$, for mock halos in the $z=0.6-0.77$ range.  The bottom panels shows a similar comparison, but for projected number density profiles.
The number density profile can be recovered better than the stellar mass profile, which could be due to the higher degree of background contamination in the stellar mass field.
To better quantify our ability to infer the true  profile shape, we rescale the radial distance by the mean virial radius of the mock clusters, and find that the \citet[][hereafter NFW]{navarro97} profile can describe both the number density and stellar mass density spatial distribution well (see the dotted curves in Figure~\ref{fig:mice2}).
Assuming that both the mean mass of the clusters and the cluster center can be accurately known, it is found that we can recover the underlying radial profiles well, in the sense that the concentration of the NFW profiles can be estimated to within 20\%.

%%%%%%%%%%%%%%%%%%%%%%%%%%%%%%%%%%%%%%%%%%%%%
\subsection{Top N Cluster Selection}
\label{sec:topn}

The redshift bins used in our analysis are at $z=0.3-0.6$, $0.6-0.77$, $0.77-0.9$, $0.9-1.02$, each occupying about a comoving volume of $(423.6\,h^{-1}\,{\rm Mpc})^3$.
Extending our previous work in \citet{inagaki15}, we employ the {\em top N} cluster selection in each of these redshift bins, and posit that these clusters could be regarded to represent a progenitor-descendant relationship.  
Underpinning the usefulness of the {\it top N} selection are the assumptions that (1) the rate of mass assembly is similar in massive galaxy clusters and (2) the merging rates among massive clusters are negligible.  As such, the top $N$ most massive clusters at a given redshift would remain the top $N$ most massive clusters at a later epoch, thus naturally providing a progenitor-descendant relationship. 
While extensive tests have been presented in \citet{inagaki15},
here we use data from the Millennium Run simulation \citep{springel05} to further test the validity and limits of this approach, concluding that even though these assumptions are not strictly true, the {\it top N} selection remains a very useful approach for studying evolution of clusters and their associated galaxy populations.

We use the version of the simulation run with the {\it WMAP7} cosmology for our tests, and only consider a cubical volume with $424\,h^{-1}\,{\rm Mpc}$ on a side.  Four snapshots, at $z=0.45, 0.68, 0.83$, and $0.98$, are considered, and we use merger trees to figure out the progenitor-descendants of the halos in these snapshots.
In the perfect case where the halos are selected by their mass, we find that between $z=0.98$ and $0.83$, 86\% of the top $N=100$ halos selected at the higher redshift remains among the top 100 at the later epoch.
The corresponding fractions for the $z=0.83 \rightarrow 0.68$ and $z=0.68 \rightarrow 0.45$ snapshots are 86\% and 79\%, respectively.  We find that, even if we consider snapshots that are further separated in time, the remaining fraction does not degrade too much: 76\% (66\%) of the top 100 halos selected at $z=0.98$ remains even at $z=0.68$ ($0.45$), which is likely due to the rarity of major mergers.  These results are summarized in Table \ref{tab:rf} (second to fourth columns).

\begin{table}
\centering
%\tabcolsep=5pt
%\caption{\footnotesize{Targets}}
\caption{Remaining Fraction (\%)}
%\vspace{1mm}
\begin{tabular}{ccccccc}
\hline
initial $z$ & \multicolumn{3}{c}{final $z$ (no scatter)} & \multicolumn{3}{c}{final $z$ (25\% scatter)}\\
 & $0.83$ & $0.68$ & $0.45$ & $0.83$ & $0.68$ & $0.45$\\
\hline
$0.98$ & 86 & 76 & 66  & 62 & 67 & 58 \\ 
$0.83$ & -- & 86 & 70   & -- & 64 & 55 \\
$0.68$ & -- & -- & 79  & -- & -- & 58\\
\hline
\label{tab:rf}
\end{tabular}
\end{table}

In reality, one cannot select clusters by their halo mass.  An observable serving as the mass proxy, such as richness or X-ray luminosity, is often used instead.  
After perturbing the true mass by a log-normal random variate with $\sigma_{\log M}=0.1$ (which is equivalent to using a proxy that exhibits a $\sim 25\%$ fractional scatter in mass,
a reasonable assumption for our richness $\hat{N}$; \citealt{oguri14,oguri17}), we find that the remaining fraction all approaches $55-70\%$, irrespective of the time interval between the snapshots.  The results are shown in Table \ref{tab:rf} (fifth to seventh columns).

As is shown in \citet{inagaki15}, the remaining fraction is not a strong function of $N$.  Given our survey volume, choosing $N=50$ or $200$ only changes the results as shown in Table \ref{tab:rf} by a few percent.  We adopt $N=100$ in this study primarily because of the desire to focus on reasonably massive clusters (e.g., $>10^{14}\,M_\odot$; see Section~\ref{sec:mass}) while still having sufficient number of clusters to work with.

\begin{figure}
\epsscale{1.25}
\plotone{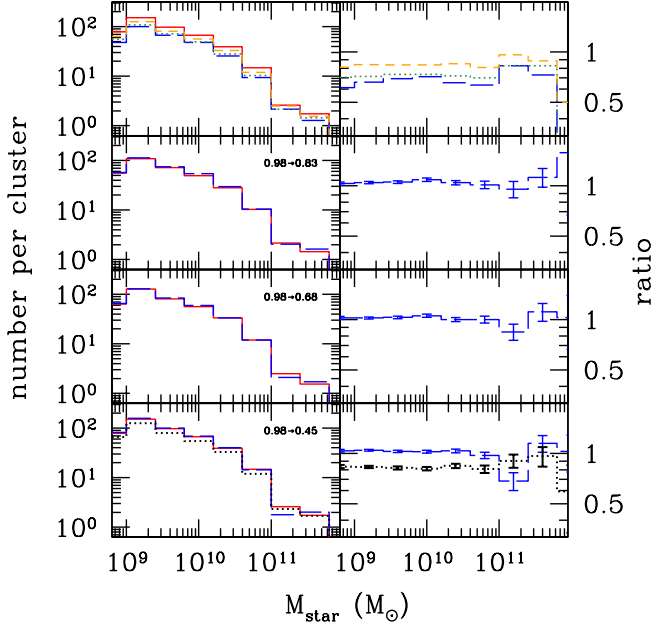}
\vspace{-5mm}
\caption{ 
The top left panel  shows the SMDs of the top 100 richest halos at $z=0.98, 0.83, 0.68,$ and $0.45$ (colored in blue, green, orange, and red, respectively) from the Millennium Simulation.  The change of the galaxy content at all masses is clear. 
In the lower three panels on the left, we show pairwise
comparisons of the SMDs between the top 100 richest halos at the three lower redshifts ($z=0.83, 0.68, 0.45$; solid histogram) and the descendant halos of the top 100 halos selected at $z=0.98$ (dashed histogram) but examined at the same lower 
redshifts, based on the semi-analytic model of \citet{guo13}.  
The four panels on the right show the ratio of non-solid histograms to the solid histogram in each of the panels on the left.
In the bottom left panel, we additionally show as dotted histogram the SMD from the top 200 halos at $z=0.45$, which shows apparent offset from the other histograms in the same panel (and can be seen more clearly in the lower right panel as the black dotted histogram).  
The facts that dashed histograms on the lower three panels on the right are all very close to unity, and that the dotted histogram is distinctly different from the dashed one in the lower right panel,
suggest that the {\it top N} cluster selection can be used to meaningfully compare the evolution of cluster galaxy population.
}
\label{fig:milsmf}
\end{figure}

For the purpose of our study,  it is more interesting to examine whether the {\em top N} selection can recover the evolution of the SMD, that is, whether the composite SMD built from the top $N$ clusters at a later epoch is representative of that of the descendants of top $N$ clusters selected at an earlier cosmic time.
For this purpose, we use
model galaxies generated by the semi-analytic model of \citet{guo13}, which is tuned to reproduce the galaxy stellar mass function and clustering in the local Universe.
From Table \ref{tab:rf}, we see that the remaining fraction ranges from 58\% to 67\% by going from $z=0.98$ to lower redshifts when a $0.1$\,dex mass scatter is introduced.  
In Figure~\ref{fig:milsmf} (lower three panels on the left) we show pairwise comparisons of the SMDs between that of the top 100 halos at $z=0.45$, $z=0.68$, $z=0.83$ (solid histogram) and that of the descendants of the top 100 halos at $z=0.98$ observed at these lower redshifts (dashed histogram), with this level of scatter in mass--observable relations.
For example, the solid histogram in the bottom panel is the SMD of the top 100 richest halos found at $z=0.45$, while the dashed histogram represents the SMD of the descendants at $z=0.45$ of the top 100 richest halos selected at $z=0.98$.
The dotted histogram, on the other hand, represent the SMD of the top 200 richest halos identified at $z=0.45$.
For ease of comparing different SMDs, in the lower right panel we show the ratios of the dashed and dotted histograms to the solid histogram.
The difference between the dotted histogram and the other two histograms in the bottom panels, together with
the good agreement between the solid and dashed histograms  in the lower three rows of the Figure (the right hand panels show the ratio of non-solid histograms to the solid histogram in the corresponding left panels), suggest that, 
even though the remaining fraction is not very high, the SMDs in clusters are universal enough (at least in the semi-analytic model) that the {\it top N} selection can still allow us to study the evolution of SMDs across  cosmic time.  
We emphasize that we have effectively assumed that the SMD depends primarily on the halo mass and only weakly on the exact assembly history of the halos.

We can also repeat the same exercise, but focusing on BCGs in these simulated clusters.  Among the BCGs in $z=0.45$ descendant halos of the top 100 halos at $z=0.98$, the median of the logarithm of stellar mass are $11.64\pm 0.02$.  
The same number for the top 100 halos selected at $z=0.45$ is $11.60 \pm 0.03$, which can be clearly distinguished from that for the top 200 halos at the same redshift ($11.54 \pm 0.02$).  
Thus it is also reasonable to expect the {\it top N} selection to provide a useful way to disentangle the stellar mass assembly history of BCGs.

%%%%%%%%%%%%%%%%%%%%%%%%%%%%%%%%%%%%%%%%%%%%%
\subsection{Cluster Mass Estimates}
\label{sec:mass}

Finally, we estimate the mean masses of our cluster samples.
As a rigorous determination of the richness--halo mass relation for {\sc camira} via weak lensing is not yet completed (although see \citealt{murata17}), we here adopt two different approaches, one from the stacked lensing and the other from an abundance consideration.  

In our lensing analysis,
the shape of each galaxy is measured on the coadded $i$-band image using the re-Gaussianization method, which measures moments of the galaxy image taking account of the non-Gaussianity of the PSF perturbatively \citep{hirata03}. Shape measurements are carefully calibrated using image simulations. We include multiplicative and additive bias derived from the image simulations in our weak lensing analysis (see \citealt{mandelbaum17} for more details).
Using the shear catalog as presented in \citet{mandelbaum17}, we
derive the average differential surface mass density profile following \citet{medezinski17}, in the range of comoving radii between $0.27h^{-1}$Mpc to $3h^{-1}$Mpc with the bin width of $\Delta(\log r)=0.15$. For photometric redshift of each source galaxy,
we adopt the {\tt mlz} photometric redshift (see \citealt{tanaka17}). As discussed in \citet{medezinski17}, the secure background galaxy selection is important for cluster weak lensing analyses. We adopt the $P(z)$ cut method in which the integrated probability density function of the photometric redshift of each galaxy is used to select galaxies behind clusters (see \citealt{medezinski17} for details). The derived differential surface mass density profile is fitted with the NFW density profile predictions to infer the average mass ($M_{200}$) and concentration parameter ($c_{\rm 200}$). 
We find that, from our highest to lowest redshift bins, the typical masses are $M_{200}=2.0, 1.9, 3.0,$ and $4.4\times 10^{14}\,M_\odot$.
The best-fit concentration parameters are $c_{\rm 200}\sim 2$, which is smaller than the expected value of $c_{\rm 200}\sim 5$ (e.g., \citealt{zhao09}). This is most likely due to the miscentering effect, which reduces the weak lensing signals near the center (although removing the innermost regions essentially has no effect in the resulting mass; see Section~\ref{sec:disc} for more discussion regarding miscentering). If we force the concentration parameter to $c_{200}=5$, the best-fit masses are decreased by $\sim 20$\%.

\begin{table*}
\centering
%\tabcolsep=5pt
%\caption{\footnotesize{Targets}}
\caption{Basic Cluster Properties}
%\vspace{1mm}
\begin{tabular}{cccccccc}
\hline
 &   &   & \multicolumn{2}{c}{stacked lensing} & \multicolumn{2}{c}{abundance}\\
bin & redshift range & mean $z$ &  $M_{200}$  &  $r_{200}$ &  $M_{200}$  &  $r_{200}$ & $\hat{N}_{\rm lim}$\\
 &  & & ($10^{14}\,M_\odot$) & (Mpc) & ($10^{14}\,M_\odot$) & (Mpc) & \\
\hline
1 & $0.30-0.60$ & 0.45 & $4.4\pm 0.2$ & 1.33 & $3.7$ & 1.27 & 30.0\\
2 & $0.60-0.77$ & 0.69 & $3.0 \pm 0.3$ & 1.07 & 3.0 & 1.09 & 22.7 \\
3 & $0.77-0.90$ & 0.84 & $1.9 \pm 0.4$ & 0.86 & 2.6 & 0.98 & 21.6 \\
4 & $0.90-1.02$ & 0.96 & $2.0 \pm 0.4$ & 0.84 & 2.1 & 0.87 & 18.0 \\
\hline
\label{tab:cls}
\end{tabular}
\end{table*}

As the second method, we make use of 
halo samples from the $(424\,h^{-1}\,{\rm Mpc})^3$
sub-volume of the Millennium simulation (see Section~\ref{sec:topn}).  
Since the exact scatter between the {\sc camira} richness and mass is still to be measured from lensing, we perturb the true mass of the halos by a set of values of $\sigma_{\log M}$ (from 0.06 to 0.16, which spans the reasonable range of the possible values), and take the average of the median mass derived from the top 100 halos for each of the $\sigma_{\log M}$ value.  From our highest to lowest redshift bins, the typical masses are found to be $M_{200}=2.1, 2.6, 3.0,$ and $3.7\times 10^{14}\,M_\odot$, in reasonable agreement with the weak lensing estimates.

\begin{figure}
\epsscale{0.9}
\plotone{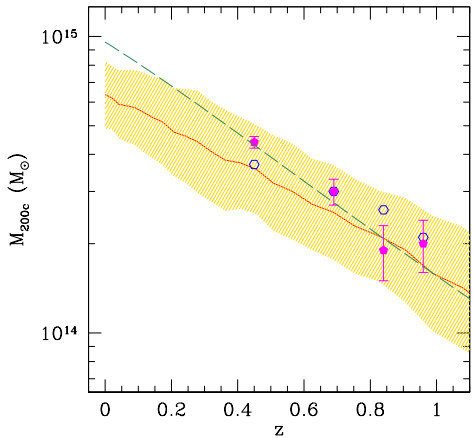}
\vspace{-2mm}
\caption{ 
The estimated mass of our cluster sample, and the probable descendant mass range.  The solid and open symbols represent the lensing- and abundance-based masses.  The solid curve and the shaded region show the median and interquartile range of the main progenitor mass of the top 100 richest halos at $z=0$ from the Millennium simulation.  The typical descendant mass at $z=0$ is $M_{200c}=6.5\times 10^{14}\,M_\odot$.  The dashed curve is the prediction from the model of \citet{zhao09}, for halos whose present-day mass is $M_{200c}= 10^{15}\,M_\odot$.
}
\label{fig:tree}
\end{figure}

In the following analysis, we adopt the weak lensing based masses as the default, but note that our results do not change if we use the abundance-based estimates.
In Table \ref{tab:cls} we provide some basic information of our cluster sample, including the redshift ranges of each of the four redshift bins, the estimated masses and radii of the sub-samples, and the limiting richness $\hat{N}_{\rm lim}$ for the top 100 selection.

We close by estimating the probable descendant cluster mass range at $z\approx 0$ for our cluster sample, making use of the mass growth history of the top 100 richest halos from (the sub-volume of) the Millennium simulation at $z=0$ (selected by assuming $\sigma_{\log M}=0.1$).
In Figure~\ref{fig:tree} we show as solid curve and shaded region the median and interquartile range of the main progenitor mass for these 100 halos.  The curve is consistent with the masses of our sample, for both lensing-based (solid symbols) and abundance-based (open points) estimates.  The median mass at $z=0$ is about $6.5\times 10^{14}\,M_\odot$.
As our lensing-based mass mainly lies in the upper part of the shaded region, we consider an independent way of estimating the descendant mass, by utilizing the mean mass growth history following the prescription of \citet{zhao09}.  The dashed curve in the Figure shows the prediction from that analytic model, which passes through the locus of the weak lensing-based mass estimates, and reaches $M_{200c}\approx 10^{15}\,M_\odot$ by $z=0$.  It is reasonable to assume the typical descendant mass lies in the range bracketed by the two methods employed here.

%%%%%%%%%%%%%%%%%%%%%%%%%%%%%%%%%%%%%%%
%%%%%%%%%%%%%%%%%%%%%%%%%%%%%%%%%%%%%%%
\section{Results}
\label{sec:res}

In Section~\ref{sec:mice} we have demonstrated that our background correction scheme works well in recovering the true SMD and the spatial distribution of galaxies, while in Section~\ref{sec:topn} it is shown that the {\it top N} cluster selection allows us to trace the cluster galaxy evolution, including BCGs, fairly accurately.  We are thus well poised to address some important topics in cluster galaxy evolution, including the stellar assembly history of BCGs, and  changes in SMD, LD,  surface stellar mass density profiles, and the radio galaxy population, as a function of cosmic time.

%%%%%%%%%%%%%%%%%%%%%%%%%%%%%%%%%%%%%%%
\subsection{Evolution of Brightest Cluster Galaxies}
\label{sec:bcg}

Strictly speaking, with our cluster sample we can only study the evolution of {\it red} BCGs, as \ca  considers solely red sequence galaxies in the cluster detection and characterization process.
Once a candidate cluster is found, a BCG is chosen to maximize the likelihood that takes into account the centrality of BCGs in the spatial distribution of member galaxies, as well as the stellar mass distribution in the massive end of the stellar mass--halo mass relation.
The cluster parameters, such as the center location and cluster redshift, are then refined iteratively, taking into consideration the properties of the candidate BCG.  The above procedure is repeated until a converged solution is found \citep{oguri14}.

\begin{figure}
\epsscale{1.5}
\plotone{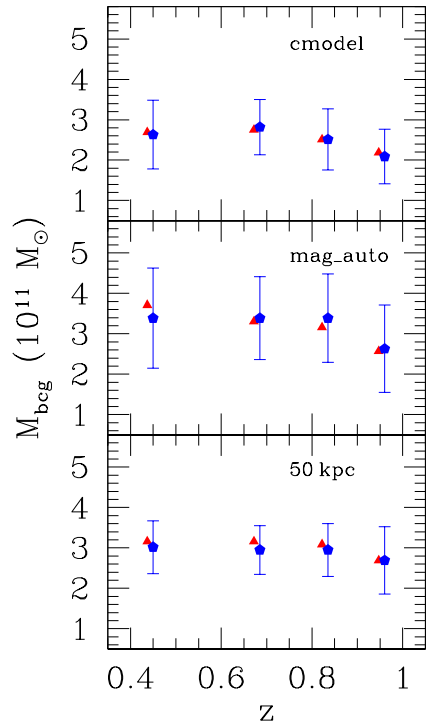}
\vspace{-8mm}
\caption{ 
Evolution of BCG stellar mass.  We consider three types of photometry to be used to infer the luminosity content of the BCGs.  From top to bottom, we show the results based on the {\tt cmodel} magnitudes as provided by the HSC pipeline, the {\tt mag\_auto} magnitude obtained using SExtractor, and an aperture magnitude corresponding to a 50\,kpc diameter, also derived from SExtractor.  
The median Kron radii, the aperture used to measure {\tt mag\_auto}, from our highest to lowest redshift bins, are 22, 28, 29, and 36 kpc, respectively.
In each panel, two methods are employed to infer the stellar mass, namely via the machine-learning algorithm {\sc demp}, and the mass provided by {\sc camira}, which is obtained as part of the cluster detection.  We show as data points the median mass of the BCGs in each of the top 100 clusters at the 4 redshift bins considered.  The blue points are for  the {\sc demp}-based mass, while the red points are {\sc camira}-based mass (scaled down by a factor of $1.7$ to account for differences in the adopted IMFs).  The error bars represent the semi-interquartile range of the BCGs; for clarity, only those for {\sc demp}-based mass are shown.
}
\label{fig:bcgsm}
\end{figure}

Our main goal here is to infer the average degree of stellar mass growth in BCGs from $z\approx 0.96$ to $z\approx 0.45$.  As BCGs are among the largest galaxies in the Universe, defining and measuring their total luminosity and stellar mass contents have always been a challenging task \citep[e.g.,][]{gonzalez05,kravtsov14}.  
As an exploratory study seeking for the first-look result on the BCG mass growth, we thus perform simplistic photometric measurements that should at least allow us to capture the luminosity in the main/inner part of the galaxies [i.e., ignoring any very extended components such as cD envelop or intracluster light (ICL)].
To really trace the light profile out to large scales (e.g., $> 100\,$kpc), one needs to carefully mask out (or model) all detectable sources on, close to, and around BCGs, and pay special attention to the sky level and scatterred light, which is left for a dedicated study in the near future.  We refer to \citet{huang17} for a detailed analysis of relatively nearby BCGs (at $z=0.3-0.5$) using the HSC-SSP survey data.

For our analysis, three measurements of the BCG flux are used.  The first one is the {\tt cmodel} magnitudes from the HSC pipeline, which are the results of a linear combination of an exponential disk model and a de Vaucouleur profile.
The {\tt cmodel} flux is the flux of the best-fit model galaxy, obtained by adding
all the profile-weighted flux in the object, so there is not a well-defined aperture
 (see \citealt{bosch17} for more details).
As mentioned above, there are still  problems with the {\tt cmodel} magnitudes from the  pipeline in crowded regions.  
Furthermore,  the {\tt cmodel} magnitudes may be systematically underestimated for relatively bright objects ($i\lesssim 20$\,mag; see \citealt{aihara17}).
We therefore run SExtractor \citep{bertin96} on the final-stacked $i$-band images and obtain the ``total'' magnitude ({\tt mag\_auto}) and aperture magnitude.  The aperture photometry is interpolated to derive the magnitude at a fixed physical  (circular) aperture, which we choose to be 50\,kpc in diameter.  
{\sc demp} is used to derive stellar mass of BCGs from the $grizy$ {\tt cmodel} magnitudes ($M_{\rm star,bcg}^{\rm cmodel}$).
Stellar masses corresponding to {\tt mag\_auto} and aperture 
magnitude are obtained by scaling from $M_{\rm star,bcg}^{\rm cmodel}$ with the  differences between $i_{\rm mag\_auto}$ and $i_{\rm cmodel}$, and $i_{\rm aper}$ and $i_{\rm cmodel}$, respectively.

Our measurements are presented in Figure~\ref{fig:bcgsm}.  
The top, middle, and bottom panels show the results using {\tt cmodel},  {\tt mag\_auto}, and the 50\,kpc diameter aperture, respectively.
In each of the panels, the blue points represent the median of the BCG stellar masses (derived with {\sc demp}) of the top 100 clusters in each of the redshift bins.
As an independent check, we also show as red triangles the stellar mass estimated by {\sc camira}.  In a nutshell, {\sc camira} uses calibrated stellar population synthesis model from BC03 with a single burst formed at $z_f=3$ and the \citet{salpeter55} IMF to derive the stellar mass of a particular galaxy by maximizing its likelihood of being on the red sequence  at a given redshift.  
After adjusting for the difference between the adopted IMFs (recall that the {\sc demp}-based masses inherit the choice of \citealt{laigle16}, that is, the Chabrier IMF), we see that 
the redshift trends indicated by the two mass estimates generally agree with each other.

We see a gentle increase in stellar mass with both {\tt cmodel} and {\tt mag\_auto} (although the increase is mostly at $z\ge 0.68$).  The growth based on the median of the former from $z\approx 0.96$ to $z\approx 0.45$ is about 25\%, while the latter suggests a $\sim 40-45\%$ increase.
On the other hand, the lack of change in the 50 kpc aperture-based stellar mass  implies the mass growth must happen primarily over larger scales, which is consistent with the findings of \citet{huang17}.  We note that the median Kron radii characterizing the {\tt mag\_auto} photometry are 22, 28, 29, and 36 kpc, respectively, from the highest to lowest redshift bins.  
The differences in stellar mass based on {\tt mag\_auto} and the 50 kpc aperture can thus be understood from the relative sizes between the Kron radii and the fixed 25 kpc radius.

\begin{figure}
\epsscale{0.9}
\plotone{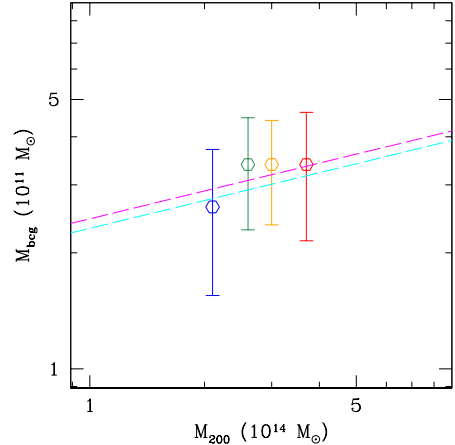}
\vspace{-2mm}
\caption{ 
Correlation between the BCG stellar mass and cluster mass.  The color of the points refers to the redshift of the clusters (from high to low: blue, green, orange, and red).  The dashed lines represent the scaling $M_{\rm bcg} \propto M_{200}^{0.24} (1+z)^{-0.19}$ from \citet{zhang16}, scaled in $M_{\rm bcg}$ so that at $z=0.45$ their relation passes through our lowest-$z$ data point.
}
\label{fig:mbmc}
\end{figure}

The degree of stellar mass growth we find here is consistent with that in \citet{lin13}, where we use a $4.5\mu$m-selected cluster sample over a $\sim 8\,$deg$^2$ area to infer a $\sim 50\%$ increase between $z=1$ and $z=0.5$. 
It is also consistent with the amount of growth seen by \citet{lidman12} and \citet{zhang16}, both of which are based on clusters detected in the X-ray.  We show in Figure~\ref{fig:mbmc} the correlation between the BCG stellar mass (based on {\tt mag\_auto}) and cluster mass (using the abundance-based estimates).  The color of the data points indicates the redshift of the clusters (from high to low: blue, green, orange, and red).
The two  dashed lines show the {\it relative} growth between $z\approx 0.96$ (cyan) and $0.45$ (magenta) from the best-fit relation of \citet{zhang16}:  $M_{\rm bcg} \propto M_{200}^{0.24} (1+z)^\gamma$ with $\gamma=-0.19$.
As our cluster sub-samples are constructed to represent clusters along an evolutionary sequence, we can clearly see  how our BCGs evolve from the lower left portion towards the upper right part in this parameter space.  Although the redshift exponent $\gamma$ in \citet{zhang16} is only weakly constrained ($-0.19\pm 0.34$), 
the consistency between our data and their relation argues for a non-negligible evolution in the BCG stellar mass--cluster mass correlation, and the stellar mass growth in BCGs.

Some of the earlier studies do not find evidence of stellar mass growth in BCGs \citep[e.g.,][]{whiley08,collins09,stott10}.
Some of these results may be due to the use of a fixed metric aperture (e.g., \citealt{whiley08}) that is reminiscent of our results using the 50 kpc aperture, while others may be attributed to the use of samples of limited sizes.
With small samples of clusters, it might be difficult to control the cluster mass distributions within the sample so that progenitor-descendant relations can be facilitated.
If the clusters at higher redshifts are selected such that they are of comparable masses as the lower redshift counterparts, then, given the seemingly weak redshift evolution of the $M_{\rm bcg}$--$M_{200}$ relation (e.g., \citealt{zhang16}; also see \citealt{brough08}), and the  large intrinsic scatter about that relation (e.g., \citealt{lin04b,lidman12}), it is plausible that subtle stellar mass growths in BCGs would be missed by previous studies that only employ small samples of clusters.

For halos of masses similar to our sample (see Table~\ref{tab:cls}), the semi-analytic model of \citet{guo13} predicts a $\sim 80\%$ increase between $z=0.98$ and $z=0.45$, which is higher compared to the  value we find.  However, given the large scatter in the mass distributions of our BCGs, this discrepancy is only at $1\sigma$ level, and is thus not significant.

%%%%%%%%%%%%%%%%%%%%%%%%%%%%%%%%%%%%%%%
\subsection{Stellar Mass Distribution}
\label{sec:smf}

We next examine the evolution of the general galaxy population in clusters, in terms of the SMD.
Following the methodology presented in Section~\ref{sec:mice}, we construct the composite SMD from galaxies within $r_{200}$ from the cluster center (using the mean $r_{200}$ for all clusters in a redshift bin), 
for each of the redshift bins.  
76 out of the 400 clusters are excluded due to large holes in the annulus or central regions, as the background estimates for these clusters would be problematic\footnote{Although in principle the effect of holes in the footprint could be circumvented by utilizing random catalogs that take into account of the ``bad'' regions in the survey, we decide not to take this route in the current analysis as we have to compute stellar mass and luminosity in the region that is used to estimate the background/foreground contribution {\it for every cluster}; such a task becomes too demanding computationally if the ``background'' region is the whole survey footprint, as will be the case when using the random catalogs.}.

\begin{figure}
\epsscale{0.9}
\plotone{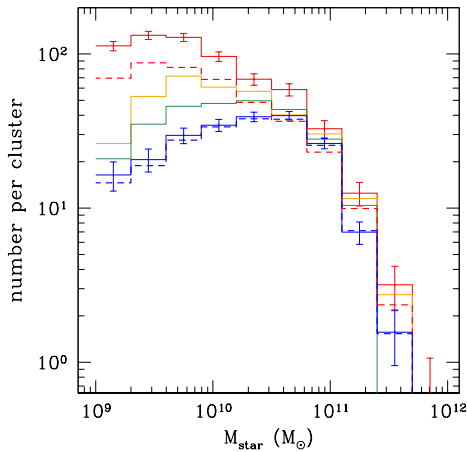}
%\vspace{-4mm}
\caption{ 
The evolution of cluster SMD (excluding BCGs).  The blue, green, orange, and red solid histograms represent the clusters from highest to lowest redshift bins.  These SMDs are measured within the mean $r_{200}$ (see Table \ref{tab:cls}). 
It is clear that the mean number of galaxies at all mass scales increases with time, particularly at both the most massive and low mass ends.
For clarity,  error bars are only shown for the SMDs of the highest and lowest redshift bins, and only account for the Poisson counting error, but not uncertainties in cluster mass.
 The blue and red dashed histograms represent the SMDs of our highest and lowest redshift cluster sub-samples, measured within a fixed 0.8\,Mpc radius.  Note that $r_{200}$ is only slightly larger than $0.8\,$Mpc for our highest-$z$ clusters.
}
\label{fig:smf08}
\end{figure}

The composite SMDs for all galaxies but the BCGs are shown in Figure~\ref{fig:smf08}.  The blue, green, orange, and red histograms are the SMDs from our highest to lowest redshift bins.  
Using the completeness curves derived in Section~\ref{sec:sample} (see Figure~\ref{fig:comp}), we have corrected for incompleteness down to $3\times 10^{9}\,M_\odot$.
We see that overall, the mean number of galaxies increases with time at all mass scales, although the effect appears to be most prominent at 
the low mass ($M_{\rm star}<10^{10}\,M_\odot$) regime, 
which is reminiscent of the findings by \citet{vulcani11}.

\begin{figure}
\epsscale{1.5}
\plotone{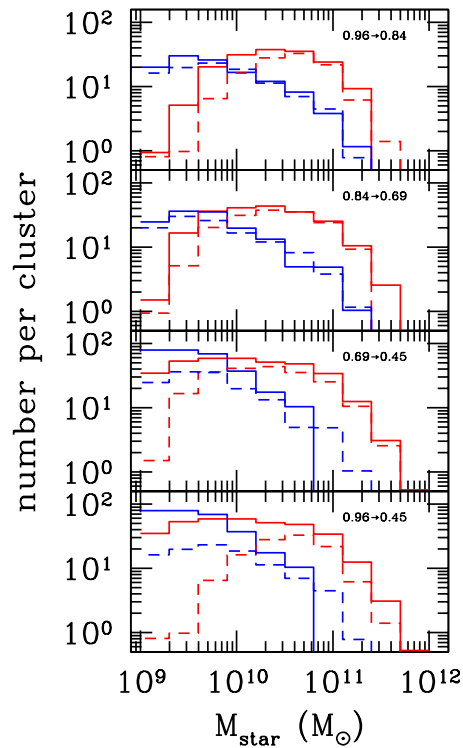}
\vspace{-8mm}
\caption{ 
Pairwise comparison of SMDs in two redshift bins, split into red and blue galaxy populations (represented by red and blue histograms).  In all panels, the solid (dashed) histogram is for the SMD of the cluster sub-sample at the lower (higher) redshift bin.  The redshift bins used for the comparison are indicated at the upper right corner in each of the panels.  These SMDs are measured within the mean $r_{200}$.
}
\label{fig:smfc08}
\end{figure}

To understand the buildup of the cluster galaxies through time, we consider the blue and red members separately.  We use the locus on color-magnitude diagrams of the red sequence as a function of redshift as provided by \ca to classify galaxies into red and blue populations. 
More specifically, as a function of magnitude, we model the observed color distribution of the red sequence galaxies by a Gaussian, and consider as red galaxies those that lie within $2.5\sigma$ from the mean of the Gaussian.  Those that are bluer are regarded as blue galaxies.  Depending on the redshift and magnitude, the typical width lies in the range $\sigma\sim 0.04-0.09$ mag.
The resulting SMDs for red and blue galaxies are shown in Figure~\ref{fig:smfc08}.  Each of the panels shows comparisons of SMDs in two redshift bins; from top to bottom, the bins considered are 0.96 {\it v.s.}~0.84, 0.84 {\it v.s.}~0.69, 0.69  {\it v.s.}~0.45, and 0.96  {\it v.s.}~0.45.
The solid (dashed) histograms represent the SMDs in the lower (higher) redshift bin, while the red and blue histograms represent the red and blue populations.
Generally speaking, for both red and blue populations, the number of galaxies increase with time at all mass scales.  At the very low mass scales (e.g., $M_{\rm star}<5\times 10^9\,M_\odot$), blue galaxies always dominate over red ones, whereas the opposite holds at the massive end (a few times $10^{10}\,M_\odot$), where the contrast becomes larger with time.
The changes are most apparent when we contrast the $z\approx 0.96$ bin with the $z\approx 0.45$ bin (bottom panel).

For these SMD measurements, we have derived the completeness curves for blue and red galaxies separately in each of the redshift bin (again using the data in the COSMOS field with the same criteria for red-blue demarcation as for the clusters), and applied completeness corrections down to $M_{\rm star}=3\times 10^9\,M_\odot$.

The abundance of red galaxies at $z\approx 0.45$ is twice as large as the number at $z\approx 0.96$ when we integrate the SMD above $10^{10}\,M_\odot$.
Given that the total number of galaxies $N$ scales with cluster mass as $N \propto M_{200}^\beta$ where $\beta\approx 0.8$, and that the scaling does not vary much with redshift (e.g., \citealt{lin04,lin06,lin12,chiu16c,hennig17}), a factor of 2 growth in number of galaxies is consistent with the expectation from the factor of 2.2 increase in cluster mass  between these two epochs.
This also implies that the change in the (red) galaxy population is mostly dominated by processes associated with cluster growth (e.g., accretion and merging with smaller galactic systems or from the field), rather than in situ star formation.

For low mass red galaxies with  $M_{\rm star}=3\times 10^9 - 10^{10}\,M_\odot$, we find that the abundance has grown by a factor of 7.3 from $z\approx 0.96$ to $z\approx 0.45$.  Comparing to the factor of 2 growth for the more massive red galaxies, we see a clear manifestation of  ``down-sizing''.  As for the blue galaxies, the relative growths over the same period for massive ($M_{\rm star}>10^{10}\,M_\odot$) and low mass ($3\times 10^9 - 10^{10}\,M_\odot$) ones are 1.5 and 3, respectively.

\begin{figure}
\epsscale{1.6}
\plotone{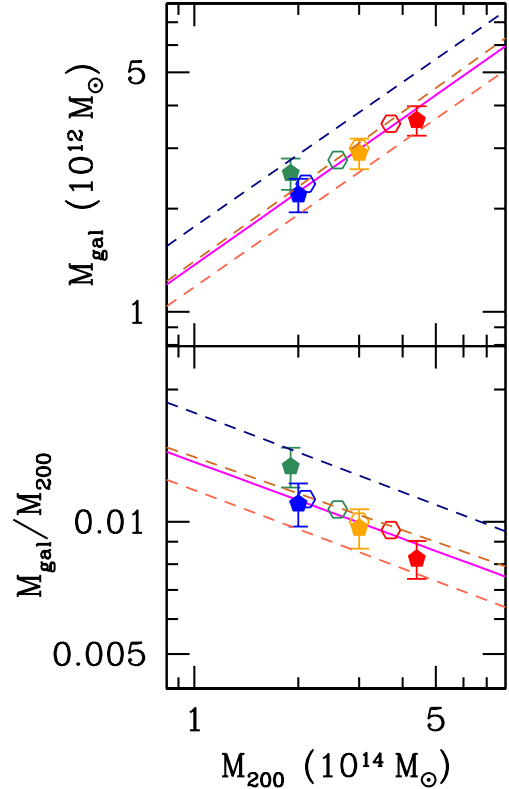}
\vspace{-8mm}
\caption{ 
The stellar mass--cluster mass correlation is shown in the top panel, while the stellar-to-total mass ratio is shown in the bottom panel. 
Only galaxies more massive than $10^{10}\,M_\odot$ are included in the $M_{\rm gal}$ measurement.
The style of the points refers to the way cluster mass is estimated (open points: based on the abundance of clusters; solid points: weak lensing; see Table~\ref{tab:cls} and Section~\ref{sec:mass}).  
The color of the points refers to the redshift of the clusters (from high to low: blue, green, orange, and red).
While clusters become more massive with time, 
the corresponding gain in stellar mass is smaller, thus causing
a decreasing stellar-to-total ratio.  The solid line is from \citet{lin12}, showing $M_{\rm gal} \propto M_{200}^{0.71}$, based on a sample of X-ray selected clusters at $z<0.6$.  Combined together, these results suggest no evolution of the stellar-to-total relation.
The three dashed lines are from our simplistic model for the stellar mass buildup (see text for  details); from top to bottom, we show the relation at $z=0.98$, 0.36, and 0.04.
}
\label{fig:fstar}
\end{figure}

We next investigate the correlation between the total stellar mass content ($M_{\rm gal}$) and the cluster mass.
Here $M_{\rm gal}$ is obtained by integrating the observed SMD down to $10^{10}\,M_\odot$,  including the contribution from the BCGs.
We find that the stellar mass content in clusters is totally dominated by red galaxies.  From our highest redshift bin to the lowest, the red galaxies account for 82\%, 82\%, 86\%, and 91\% of the  stellar mass in cluster galaxies (the corresponding numbers when BCGs are excluded are 78\%, 78\%, 82\%, and 88\%).
In the top panel of Figure~\ref{fig:fstar}, the $M_{\rm gal}$--$M_{200}$ correlation is shown, while the stellar-to-total mass ratio ($M_{\rm gal}/M_{200}$) is presented in the bottom panel. 
As we have two ways of estimating cluster mass, we present results from both methods.  In the Figure, the solid points are derived by using weak lensing-based mass, while the open ones are calculated using abundance-based mass (Table~\ref{tab:cls}). 
Both methods give consistent results.  
As in Figure~\ref{fig:mbmc}, the color indicates the redshift of the clusters (from high to low: blue, green, orange, and red).  
In Figure~\ref{fig:fstar}
we can  clearly see the direction clusters move with time; 
the more massive clusters become, the smaller the stellar-to-total ratio gets.  The solid line in the Figure is the $M_{\rm gal} \propto M_{500}^{0.71\pm 0.04}$ scaling obtained by \citet[][note that both their stellar mass and total mass are measured within $r_{500c}$]{lin12}, adjusted only in normalization  by the differences in the IMFs adopted.  It is a bit surprising to see that the line matches very well with the locus of our data points, given the differences in the two analyses (different cluster samples, redshift ranges, cluster mass definition and calibration).  Our result confirms the finding of \citet{lin12}, that the stellar mass--cluster mass relation shows no evidence for redshift evolution (see also \citealt{chiu16c}).

We try to understand the lack of redshift evolution of the $M_{\rm gal}$--$M_{200}$ relation by considering the following simplistic model, 
which may shed light on the hierarchical
buildup of stellar mass content in clusters.  
We start by constructing the complete merging history of massive halos (e.g., $M_{200}\ge 10^{14}\,M_\odot$) selected from Millennium simulation at $z=0$, including progenitor halos down to galactic scale halos ($M_{200}\ge 10^{11}\,M_\odot$).  Whenever a halo is formed, a stellar mass is ``assigned'' to it, following a certain $M_{\rm gal,ini}(M_{200})$ function that is assumed to be invariant in time.  When a halo merges with a more massive halo, 
a fraction $f_{\rm loss}$ of stellar mass is assumed to be lost to the intrahalo or interhalo space (thus becoming unaccounted for with our SMD measurements).  
Our goal is to see if the model can be tuned to reproduce the observed behavior of the $M_{\rm gal}$--$M_{200}$ relation, that is, with no noticeable evolution in both amplitude and slope.

For simplicity, we start tracking the stellar mass buildup at $z=3$, and assume the $M_{\rm gal,ini}(M_{200})$ function to be a power-law ($M_{\rm gal,ini} = A\, M_{200}^\theta$).  With such a setting, there are two relevant parameters in our model, namely the stellar mass loss fraction $f_{\rm loss}$ and the power-law index $\theta$.
Without stellar mass loss (i.e., $f_{\rm loss}=0$), the slope of the resulting $M_{\rm gal}$--$M_{200}$ relation is steeper than the observed value of $\approx 0.7$, irrespective of the value of $\theta$.  
Some stellar mass loss is thus required to balance the accumulation of stellar mass in massive halos.
A set of combination of these parameters is found to 
reproduce the value of the slope of the $M_{\rm gal}$--$M_{200}$ relation and its lack of redshift evolution, although the model still results in weak evolution in the amplitude (which may be due to mass accretion below the mass limit in our merger tree treatment, and/or mergers taking place in between simulation snapshots).  
Without an exhaustive exploration of the parameter space, we find that a model with $(f_{\rm loss}, \theta)=(0.4, 0.5)$ appears to work well (see the dashed lines in Figure~\ref{fig:fstar}).  
The best model also suggests that about $10-15\%$ of the final stellar mass is from the initial assignment to the main progenitor, about equal amount ($\sim 23\%$) from major and minor mergers (mass ratios of $\lesssim 3$ and $\lesssim 20$ respectively), and the rest ($\sim 40\%$) from accretion of smaller systems; a large contribution from the small galactic systems is consistent with the conclusions of \citet{chiu17}.

Admittedly the model is rather crude, but it does show that a quasi-steady state could be obtained (in terms of the slope of the stellar mass--cluster mass relation), and it allows us to estimate the relative contribution to the stellar mass content from progenitors of different masses.  With more detailed treatments in both the stellar mass loss process and the $M_{\rm gal,ini}(M_{200})$ function, such as variation with time and halo mass dependence,  the model might be tuned to generate the non-evolving $M_{\rm gal}$--$M_{200}$ relation as observed, but it is entering the regime of semi-analytic modeling and is left for a future study.

\begin{figure}
\epsscale{1.5}
\plotone{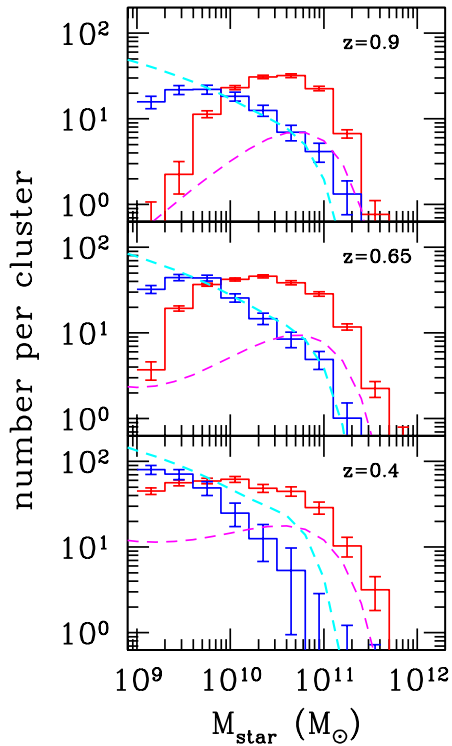}
\vspace{-8mm}
\caption{ 
Comparison between the cluster and field SMDs, for red and blue galaxies (red and blue histograms, respectively).  From top to bottom panels, we show as histograms the SMDs in clusters at $z=0.8-1$, $z=0.5-0.8$, and $z=0.3-0.5$; the redshift binning is different from our default, as we need to match the field measurements from the COSMOS survey \citep{davidzon17}.  
The error bars again only account for the Poisson error.
The field stellar mass functions (magenta and cyan dashed curves) have been scaled in amplitude to account for differences in densities in the field and cluster environments.
}
\label{fig:cfcosmos}
\end{figure}

Finally, it is instructive to compare our measurements with the stellar mass function in the field, 
taken from the latest study using data from the COSMOS survey \citep{davidzon17}.  As the COSMOS study employs the restframe $UVJ$ color-color diagram for the distinction between red and blue galaxies, while we use the red sequence color in optical bands, we do not expect the comparison to be exact (for example, we cannot distinguish  quiescent galaxies from very dusty ones).
Since the redshift bins used by \citet{davidzon17} are  different from what we have adopted ($z=0.2-0.5$, $0.5-0.8$, $0.8-1.1$), we have re-grouped our clusters into three redshift bins ($z=0.3-0.5$, $0.5-0.8$, $0.8-1.02$) and re-measured the SMDs.  In Figure~\ref{fig:cfcosmos} we show the results of this comparison.  The  histograms show our cluster SMDs; the mean redshift is shown in each of the panels.  The dashed curves are the \citet{schechter76} function fits to the observed stellar mass functions from COSMOS, scaled in amplitude by a factor of $(4\pi/3) r_{200}^3 \times 200/\Omega_M(z)$ to account for differences in mean densities of the two environments.
We find generally a good agreement in the overall shape for both red and blue populations.  
We see that the cluster red galaxy abundance is always much higher compared to the (scaled) field value, indicating that the red cluster galaxy population is not simply produced by the agglomeration of the quenched field population; the cluster environment (or any ``pre-processing'' accompanying the hierarchical cluster growth) must have much enhanced the quenching processes.
On the other hand, the good agreement in amplitude of the blue SMDs  (except for the lowest-$z$ bin) suggests that clusters at $z>0.5$ still have a ``fair'' share of the blue galaxy population.

The  difference in amplitude between the cluster and scaled field SMDs of blue galaxies in the $z=0.3-0.5$ bin, together with the overabundance of massive, red galaxies with respect to the field (particularly at the highest redshift bin), may be regarded as a manifestation of the environmental dependence of the down-sizing phenomenon \citep{tanaka05};  while the most massive galaxies ($M_{\rm star}\sim 10^{11}\,M_\odot$) in clusters have already been quenched and become red at $z\gtrsim 0.9$, quenching has finally progressed to lower mass cluster galaxies (a few times $10^{10}\,M_\odot$) by $z\sim 0.4$, thus lowering the abundance of blue cluster galaxies with respect to the expectations from the field (see also \citealt{vulcani13,vanderburg13}).

%%%%%%%%%%%%%%%%%%%%%%%%%%%%%%%%%%%%%%%
\subsection{Luminosity Distribution}

\begin{figure}
\epsscale{1.32}
\plotone{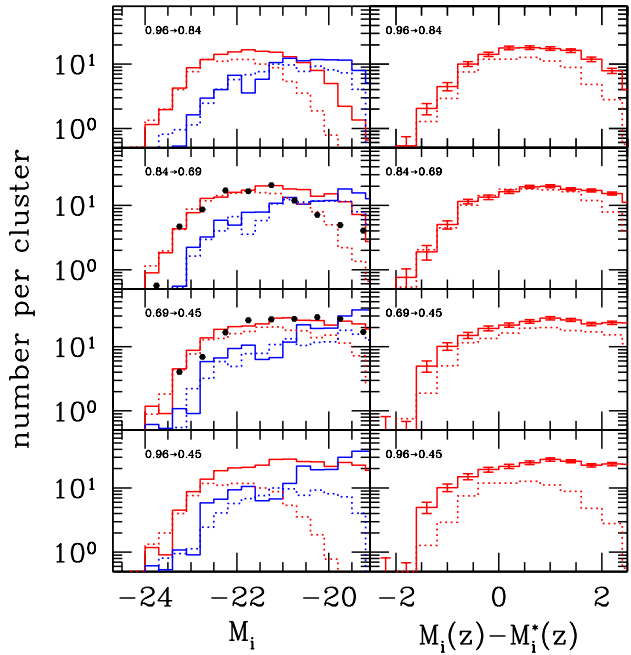}
\vspace{-7mm}
\caption{ 
The evolution of the cluster LD within $r_{200}$.  The left panels are similar to Figure~\ref{fig:smfc08}, showing pairwise comparisons of LDs in different redshift bins.  
The right panels show only the LDs of red galaxies, after passive evolution has been removed.
The black points in two of the panels on the left represent the LDs measured by the EDisCS survey \citep{rudnick09}.  The points on the second panel from the top (bottom) are the LD for their clusters at $z=0.6-0.8$ ($0.4-0.6$), which can be compared to the red solid histogram in the same panel.  
For clarity, error bars are only shown for the solid histograms on the right hand panels, and only account for the Poisson error.
}
\label{fig:lf}
\end{figure}

Another fundamental statistic characterizing the galaxy population is the LD. 
The composite LDs are constructed following the same methodology as for the SMDs, and are shown on the left panels of Figure~\ref{fig:lf}.  
Completeness corrections, derived in an analogous way to those used for the SMDs, have been applied.
Consistent with the finding from the previous section, the red galaxies dominate over blue ones among the luminous members.  
For the low luminosity galaxies, the increase of the red population is very dramatic, which is consistent with the finding of \citet[][see below]{delucia07b}.

\begin{figure}
\epsscale{0.9}
\plotone{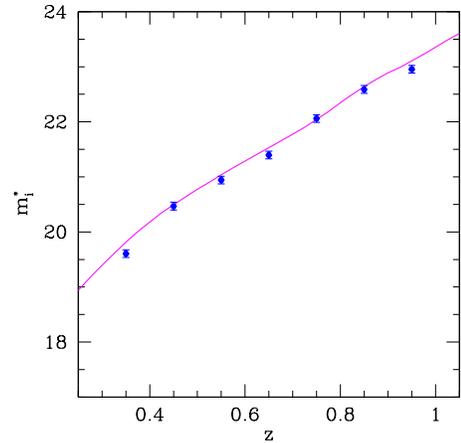}
%\vspace{-4mm}
\caption{ 
The evolution of the characteristic magnitude (in apparent magnitude in the $i$-band), measured from LDs of red galaxies constructed in fine redshift bins.  The curve shows a single burst model predicted by the population synthesis model of \citet{conroy09}, formed at $z_f=3.5$ with the Chabrier IMF.  This model is used to correct for passive evolution for the LDs shown on the right panels in Figure~\ref{fig:lf}.  See the last column in Table~\ref{tab:absm} for the absolute magnitudes from this model at the mean redshifts of our sub-samples.
}
\label{fig:apmstar}
\end{figure}

One can see that the LD of higher redshift clusters often shows a (slightly) higher amplitude in the bright end compared to that of the lower redshift LDs.  This is likely due to the passive evolution of stellar populations.
To correct for this effect and reveal the true level of evolution, 
we seek for simple stellar population models that  best describes our data.  We start by constructing composite LDs of red galaxies in the apparent magnitude space in fine redshift bins, following the methodology presented in \citet{lin12}.  The best fit characteristic magnitude ($m^{\star}_i$), derived by fitting a \citet{schechter76} function to the observed LDs, as a function of redshift is shown in Figure~\ref{fig:apmstar}.
After comparing the predictions from single burst models constructed from combinations of various stellar population synthesis models (BC03; \citealt{maraston05, conroy09}) and different formation time $z_f$ and IMFs, we find that the \citet{conroy09} model with $z_f=3.5$ and Chabrier IMF fits our observations best.  This model is shown as the curve in Figure~\ref{fig:apmstar}.

We then construct composite LDs in the absolute magnitude space again (for red galaxies only), but this time subtracting all the magnitudes by the evolving absolute magnitude of the single burst model, effectively taking out passive evolution.  The results are shown in the right panels of Figure~\ref{fig:lf}.
After the passive evolution has been removed, 
we can see the true level of growth with time, which is similar to that found from the SMDs.

We show as black points in two of the panels on the left side of Figure~\ref{fig:lf} the LDs measured by the EDisCS survey \citep{rudnick09}.  The points on the second panel from the top (bottom) are the LD for their clusters at $z=0.6-0.8$ ($0.4-0.6$), which can be compared to the red solid histogram in the same panel.  
We see that while at $z\approx 0.45$ the two measurements agree, their $z\approx 0.7$ LD shows a much more dramatic downturn at the faint magnitudes than ours, and is actually in better agreement with our $z\approx 0.84$ LD (dotted histogram in the second panel from top).  It is unclear what the cause of the discrepancy is, but our measurement suggests the emergence of faint red galaxies occurs over a longer period of time compared to what EDisCS has shown.

\begin{table}
\centering
\caption{$i$-band Absolute Magnitude $v.s.$ Stellar Mass}
\begin{tabular}{cccccc}
\hline
$z$ & \multicolumn{2}{c}{$10^{11}\,M_\odot$} & \multicolumn{2}{c}{$10^{10}\,M_\odot$} & $M^\star_i$\\
      &  red  &  blue & red & blue & \\
\hline
$0.45$ & $-22.43$ & $-22.76$ & $-20.68$ & $-21.07$ & $-21.91$\\
$0.68$ & $-22.57$ & $-23.02$ & $-20.75$ & $-21.19$ & $-22.16$ \\
$0.83$ & $-22.77$ & $-23.04$ & $-20.81$ & $-21.56$ & $-22.29$ \\
$0.96$ & $-22.91$ & $-23.04$ & $-20.85$ & $-21.50$ & $-22.39$ \\ 
\hline
\label{tab:absm}
\end{tabular}
\end{table}

We finish by providing a look-up table that allows one to check the correspondence between the $i$-band absolute magnitude ($M_i$) and stellar mass, based on the \citet{laigle16} catalog (and with our definition of red and blue galaxies).  In Table~\ref{tab:absm} we show, at 4 redshifts, the median $M_i$ corresponding to $M_{\rm star}=10^{10}\,M_\odot$ and $10^{11}\,M_\odot$, for blue and red galaxies separately and, in the last column, the absolute magnitude of the evolving characteristic magnitude ($M^\star_i$) from the best-fit passive evolution model (corresponding to the curve shown in Figure~\ref{fig:apmstar}).
Down to $M_{\rm star}=10^{10}\,M_\odot$, we are probing to $M^\star_i+1.5$ ($M^\star_i+1.2$) for red galaxies at our highest (lowest) redshift bin.

%%%%%%%%%%%%%%%%%%%%%%%%%%%%%%%%%%%%%%%
\subsection{Surface Mass Density Profile}
\label{sec:mprof}

\begin{figure}
\epsscale{1.2}
\plotone{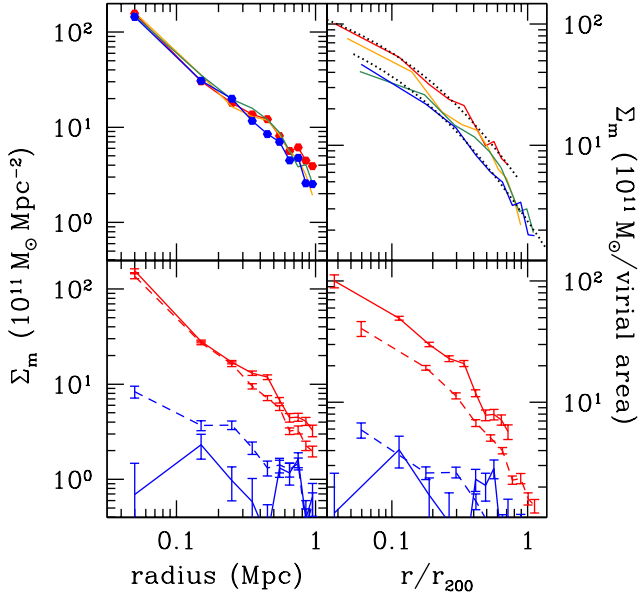}
\vspace{-8mm}
\caption{ 
Evolution of the stellar mass surface density profiles.  
The left hand panels show the profiles for which the distance from the cluster center is in unit of Mpc, while in the right hand panels the distance is scaled by the mean $r_{200}$ in each redshift bin.
The contribution from the BCGs is included (excluded) in the left (right) hand panels.
In the top panels, the blue, green, orange, and red curves show the evolution from the highest to lowest redshift bins.  For the highest and lowest bins in the top left panel, data points are also shown.
In the top right panel,
for  clusters at $z\approx 0.45$ (0.96), we show an NFW profile with $c=2.7^{+0.3}_{-0.2}$ ($3.1^{+0.3}_{-0.2}$), which describes the measurement well.
In the two bottom panels, only the profiles from the highest and lowest redshift bins are shown (dashed and solid, respectively).  The red and blue curves denote the contribution from red and blue populations.
}
\label{fig:mprof2p}
\end{figure}

It is interesting to investigate how the stellar mass content is built up spatially with time.
We thus study  the evolution of the spatial distribution of stellar mass in clusters.
The averaged profiles, produced by our statistical background subtraction method as outlined in Section~\ref{sec:mice}, are shown in Figure~\ref{fig:mprof2p}.  
We include only galaxies more massive than $10^{10}\,M_\odot$.
In the left panels, the projected distance from the cluster center is in unit of Mpc, while that in the right panels is scaled by the mean $r_{200}$.
In the top two panels, the blue, green, orange, and red curves are the profiles from our highest to lowest redshift bins.  
For clarify, data points are only shown for the profiles of clusters in the highest and lowest bins in the top left panel.   
The contribution from the BCGs is included in the left panels, but not in the right hand ones; the difference is only in the innermost radial bin.
In the bottom two panels, only the results for the highest and lowest redshift bins are shown (dashed curve for the highest, solid one for the lowest), now split by color (red curves are for red galaxies, while blue curves are for the blue ones).

It is apparent from the top left panel that the contribution from BCGs makes the profile rather peaky, uncharacteristic of the NFW profile that is found to describe the spatial distribution of (non-BCG member) galaxies in clusters \citep[e.g.,][]{lin04,hennig17}.  From the bottom left panel we see that blue galaxies do not play a significant role in the stellar mass density at any redshift, and thus the growth of the surface mass density is mainly driven by the red galaxy population, which is consistent with the finding of  \citet{vanderburg15}.  
By comparing the mean profile from $z\sim 1$ clusters with that of $z\sim 0.15$ clusters, \citet{vanderburg15} observe an ``inside-out'' growth of stellar mass density, in the sense that high-$z$ clusters show an excess (deficit) of stellar mass in the inner (outer) parts compared to the low-$z$ ones.  This is not found in our comparison of $z\approx 0.96$ and $z\approx 0.45$ clusters (dashed and solid red curves in the lower left panel), from which we see that the profile of the lower-$z$ clusters is always above that of the higher-$z$ one.  
We note that the $z\sim 1$ clusters studied by  \citet{vanderburg15} are of masses comparable to ours at $z\approx 0.96$, and that both studies employ schemes to ensure clusters along evolutionary sequences are traced.
A late buildup of ICL at $z\lesssim 0.4$ may be able to reconcile the two results.
Another possibility is a potentially higher fraction of our clusters with mis-identified center at higher redshifts, although this does not seem likely based on our tests using \ca clusters with X-ray counterparts (see Section~\ref{sec:disc}).

From the top right panel it can be seen that the stellar mass density increases with time at all radii, when the growth in cluster size is taken into account (i.e., when radius is scaled by the mean $r_{200}$).  This is true also when BCGs are included (not shown).  
Excluding the BCGs allows us to compare the stellar mass profiles with the NFW one.
We find that the $z\approx 0.45$ profile can be well fit by an NFW profile with $c=2.7^{+0.3}_{-0.2}$, while that at $z\approx 0.96$ can be described by one with  $c=3.1^{+0.3}_{-0.2}$.  These profiles are shown as dotted curves in the top right panel.
Given the uncertainties in the fitting, there is no strong evidence suggesting evolution of the profile shape between $z\approx 0.96$ and $z\approx 0.45$.

We conclude this section by commenting on the shape of the  stellar mass surface density profiles for the red and blue galaxies.  From the measurements shown in the lower right panel, we find that at $z\approx 0.96$, the red and blue galaxies can be described by an NFW profile with $c=3.1$ and $c\approx 1$, respectively.  At $z\approx 0.45$, while the red galaxies follow an NFW profile with $c=2.7$, the blue profile is extremely noisy and, if anything, would be described by a flattened/cored profile.  
Thus, 
we would conclude 
that the fraction of red galaxies always increases towards the cluster center (see also \citealt{goto04,hansen09} and references therein), and this trend becomes stronger towards lower redshift.

%%%%%%%%%%%%%%%%%%%%%%%%%%%%%%%%%%%%%%%
\subsection{Radio Galaxies}

Finally, we study the population of radio-active galaxies in clusters.  We first construct the 1.4 GHz RLD by cross correlating our cluster sample with the radio source catalog from the FIRST survey, then study the fraction of galaxies that are radio active by cross matching the FIRST sources with the HSC photometric objects in and around the clusters.  To enable comparison with previous studies, we measure  the RLD  within the estimated mean virial radius $r_{200}$.

\begin{figure}
\epsscale{0.9}
\plotone{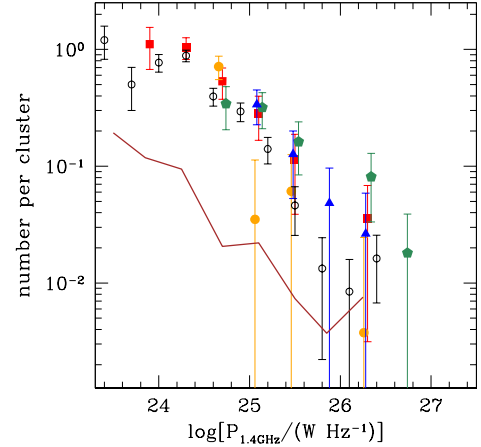}
%\vspace{-4mm}
\caption{ 
The cluster RLDs at 1.4 GHz, in unit of number of galaxies per cluster per dex in radio luminosity, measured within the estimated typical $r_{200}$.  The blue, green, orange, and red points represent RLDs from highest to lowest redshift bins.  Note that the RLDs at different redshifts are probed to different luminosity limits because of a fixed 1\,mJy flux limit.  The black points represent the $z<0.2$ RLD based on a large sample of X-ray selected clusters, taken from \citet{lin07}.  The brown curve shows the field radio luminosity function measured in the COSMOS field \citep{smolcic09}, scaled in amplitude to account for differences in density between the two environments.
}
\label{fig:rlf}
\end{figure}

The RLDs are constructed in a similar fashion as the SMDs.  
For a given cluster, we regard all radio sources with a flux greater than 1\,mJy around it to be at the cluster redshift; counts  within $r_{\rm cl}=r_{200}$ and in a large background region (of about $1500\,$deg$^2$) are recorded as a function of radio luminosity $P_{1.4}$ (a power-law index of $-0.8$ is assumed in the conversion from flux to luminosity).  The contribution from all the clusters  are summed and then the expected background level is subtracted.  
The resulting RLDs are shown in Figure~\ref{fig:rlf}.
Because of the fixed flux limit, the RLDs at different redshifts are probed to different luminosity limits.
Given the large error bars, it is hard to determine the exact shape of the RLD and its redshift evolution, although there is some hint of higher abundance of high power sources in the higher redshift bins (e.g., $z>0.77$).
The black points in the Figure are from \citet{lin07}, representing the RLD of $z<0.2$ clusters, scaled by the typical volume of clusters used in that study.  The amplitude of this local RLD is similar to, if slightly lower than, our $z\approx 0.45$ RLD, hinting at some evolution between $z<0.2$ and $z\approx 0.45$.

While there have been several studies of the redshift evolution of RLDs (or radio luminosity functions) in clusters \citep{sommer11,ma13,gupta17}, we note that our measurements are the first one probing beyond $z=0.7$.
However, given the rarity of the radio galaxies, a much larger cluster sample is needed to tightly constrain the shape and possible evolution of the RLD.

Although it would be informative to compare with radio luminosity function in the field, we note that most of such studies are based on pencil beam surveys, and thus cannot probe the most powerful radio sources.  
An interesting comparison we can make is regarding
the ``overdensity'' of radio galaxies in clusters with respect to the field.  The solid curve in Figure~\ref{fig:rlf} is the field radio luminosity function from \citet{smolcic09} at $z=0.9-1.3$, scaled in amplitude to facilitate the comparison with cluster RLDs by $(4\pi/3) r_{200}^3 \times 200/\Omega_M(z)$, as has been done for a similar comparison on SMDs in Section~\ref{sec:smf} (we have used $r_{200}=0.8\,$Mpc for this rough estimate).  To match the curve to the RLD of our $z=0.77-0.9$ or $z=0.9-1.02$ bin, we need to further scale it up by a factor of 10.  This exercise shows that even at high-$z$, when the whole Universe is expected to be more active, the cluster environment further galvanizes radio activities.  We note that for $z<0.2$, \citet{lin07} estimate that the enhancement of radio activity in clusters is a factor of $\sim 7$, thus hinting on an even stronger promotion of dense environment on radio activity at $z\gtrsim 1$.

Next we investigate the fraction of galaxies that have $P_{1.4}$ above a threshold $P_{\rm lim}$ chosen to be high enough that the origin of the radio activity is certain to be due to the central super massive black hole.
We measure the ``radio active fraction'' (RAF) again using a statistical background correction method; in both the cluster region ($r\le r_{\rm cl}$) and the annulus region, we measure the numbers of galaxies in  a given stellar mass bin and with $P_{1.4} \ge P_{\rm lim}$.  The RAF is then the ratio of the number of radio galaxies to the number of galaxies in the stellar mass bin, both corrected for the background value estimated from the annulus region.  We accumulate the numbers of both radio active and quiet galaxies from all clusters, and subtract the total background contribution before calculating the averaged RAF.  In practice, we choose $\log P_{\rm lim}=24.7$, $r_{\rm cl}=r_{200}$.

\begin{figure}
\epsscale{0.9}
\plotone{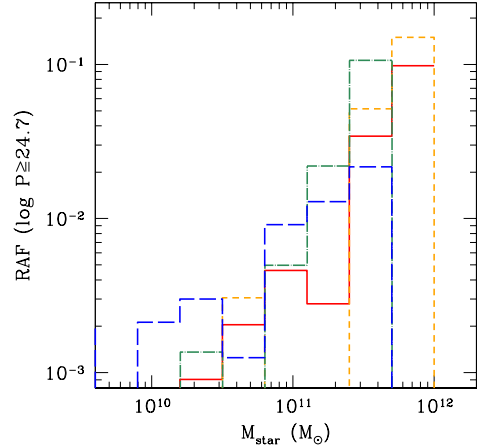}
%\vspace{-4mm}
\caption{ 
The radio active fraction as a function of stellar mass, for galaxies within the estimated $r_{200}$ from the cluster center.  The blue, green, orange, and red histograms show the results from highest to lowest redshift bins.  We see that the RAF is a strong function of stellar mass, and shows a nontrivial redshift dependence.  
Considering all galaxies more massive than $10^{10}\,M_\odot$, the RAF in the two higher redshift bins is $1.5-2$ times higher than that in the two lower redshift bins.
}
\label{fig:raf}
\end{figure}

The resulting  RAF as a function of stellar mass  is shown in Figure~\ref{fig:raf}.  We see that it is a strong function of stellar mass, which is consistent with the findings at low-$z$ \citep[e.g.,][]{best05b}.
There is a nontrivial dependence on redshift, however.
Considering all galaxies more massive than $10^{10}\,M_\odot$, the RAF in our two highest redshift bins is $1.5-2$ times higher than that of the two lower redshift bins.
It is worth noting that at $z>0.77$, the high RAF  for galaxies more massive than $10^{11}\,M_\odot$ is mainly due to blue galaxies.  Although such massive blue population disappears at lower-$z$ (see~Figure~\ref{fig:smfc08}), the appearance of massive red galaxies increases the RAF in the two lower redshift bins.
We may thus have witnessed a likely change in the dominant accretion mode powering radio galaxies in clusters at $z\sim 0.8$ or so (e.g., from cold gas-powered high Eddington ratio mode to low Eddington ratio mode).

Finally, we note that the RAF of BCGs is found to be about 7\%, which is also consistent with the value found by \citet[][c.f.~Table 5 therein]{lin07}.  Interestingly, it does not show much redshift dependence, which is consistent with the finding of \citet{gralla11}.

%%%%%%%%%%%%%%%%%%%%%%%%%%%%%%%%%%%%%%%
%%%%%%%%%%%%%%%%%%%%%%%%%%%%%%%%%%%%%%%
\section{Discussion}
\label{sec:disc}

With nearly 2000 clusters over $230\,$deg$^2$, the parent cluster sample used for the current analysis is already the largest published cluster sample over the redshift range $z=0.1-1.1$.
The full HSC cluster sample, to be realized with the complete survey over $1400\,$deg$^2$ and with better spectroscopic calibration of the stellar population synthesis models, together with weak lensing cluster mass calibration, would offer an unprecedented opportunity to improve upon what we have presented here (see also \citealt{jian17} and \citealt{nishizawa17}), allowing for extremely detailed studies of the infall, star formation, quenching, and merging of cluster galaxies out to $z\sim 1.4$ or so, with vanishingly small statistical uncertainties.  Over the HSC-deep fields, one can utilize the available near-IR data and follow the cluster evolution to much higher redshifts.

We shall continue to improve all aspects of our analysis, particularly on the identification of BCGs, the centroiding of clusters, and the way to link progenitor clusters with descendants (in a statistical sense).  We describe possible ways forward for each of these in turn.

It has been noted that as redshift increases, the star formation activity in BCGs generally rises \citep{mcdonald16}.  It is thus entirely probable that a strictly red sequence based cluster detection algorithm would not be able to identify the true BCG in a cluster whose BCG is forming stars.  Indeed,  the well-established cluster finder  redMaPPer \citep{rykoff14} does miss out the BCG in some of the low-$z$ strong cool-core clusters.
At $z\sim 1$, when 30\% of the BCGs may be star-forming  \citep[e.g.,][]{mcdonald16}, it is thus necessary to supplement \ca with an algorithm that includes blue galaxies for the consideration of BCG candidates.

A related issue is the correct identification of the true BCG, particularly towards $z\sim 1$, even if the candidates are all quiescent.  From ground-based imaging, even with the quality data from the HSC-SSP survey, often times even experienced observers cannot unanimously agree on the choice of the BCG.  It may thus be desirable to consider all probable BCG candidates, taking into account the likelihood to be the true BCG among the candidates, when studying the BCG assembly history\footnote{We acknowledge M.~McDonald for this idea.}.  We note that redMaPPer does provide a list of probable BCGs, a feature that could also be implemented in {\sc camira}.

By cross matching \ca clusters with clusters detected in the X-rays from the XXL and XMM-LSS surveys \citep[][]{pierre04,pierre16}, \citet{oguri17} find that about 30\% of the \ca BCGs are significantly offset from the X-ray emission peak.  Assuming that the X-ray peak represents the bottom of the gravitational potential of a cluster (and thus is a good proxy of the cluster center), and that the true BCG should lie close to the center \citep[e.g.,][]{lin04b,song12}, this implies that about a third of our BCGs may be mis-identified (which may contribute to the large scatter in BCG luminosity/stellar mass as seen in Figure~\ref{fig:bcgsm}).
Without extensive spectroscopy of a representative sample of our clusters, it is hard to estimate the effect of the mis-identification on the BCG stellar mass growth history inferred in Section~\ref{sec:bcg}, but we note that the fraction of \ca clusters with large offsets between the BCG position and the X-ray centroid does not vary with redshift.  Therefore, as long as the mis-identified BCGs only add noise to the true BCG stellar mass distributions (as would be the case if the mis-identified BCGs are primarily foreground/background objects), the mild growth we infer should remain robust.
In the case that \ca somehow tends to select other member galaxies (e.g., the second brightest cluster galaxy, G2) as the BCG, however, the interpretation of our results may then depend on the evolution of the magnitude gap between the BCG and G2 (or more generally the satellite SMD).  In a scenario where the typical magnitude gap increases with time (i.e., larger at lower-$z$), the slow stellar mass growth seen in Figure~\ref{fig:bcgsm} could be caused by the mis-identification, and we would in fact underestimate the true growth.  We shall return to this issue in a future study.

Conventionally, for optically selected clusters, the position of the BCG is taken as the center of a cluster (e.g., \citealt{koester07}).  There are also arguments for using the BCG location even when X-ray centroid is available (\citealt{george12}).
The large fraction of mis-identified BCGs in our current cluster sample also impacts our measurements of the projected stellar mass density profile, as we take BCGs as the cluster center.
In the future, we could use the luminosity or stellar mass weighted mean position of member galaxies as an  alternative for the cluster center. 
The best choice for the cluster center could be the one that maximizes the stacked lensing signal around the clusters,
following the approach of \citet{george12}.

We have shown that the {\it top N} selection is a promising way of constructing cluster samples at different cosmic epochs that may represent a progenitor-descendant relationship.
One possible way to refine this approach is to consider the growth mode of clusters, e.g., whether a cluster sample is fast or slow growing, which may be identified by its mean splashback radius \citep[e.g.,][]{diemer14}.

In the current study we have set a lower limit in redshift at $z=0.3$.  The primary reason for not extending the sample to lower-$z$ clusters is the limited comoving volume at $z<0.3$ (that is, for a given solid angle, the comoving volume between $z=0$ and 0.3 is much smaller than that occupied by each of our redshift bins).  Upon the completion of the HSC-SSP survey, the comoving volume between $z=0$ and 0.3 over $1400\,$deg$^2$  would be similar to that used in each of the redshift bins in the current analysis, and thus we can  study cluster evolution with the {\it top N} selection from $z=0$ all the way to  $z\approx 1.4$ (albeit using smaller solid angle for $z>0.3$).
Alternatively, we can complement our HSC sample with $z<0.3$ \ca clusters from SDSS \citep{oguri14}, or modify the redshift binning with reduced  comoving volume in each  bin.

With a much larger cluster sample with weak lensing calibrated masses, we would be able to study various scaling relations in addition to the $M_{\rm bcg}$--$M_{200}$ and $M_{\rm gal}$--$M_{200}$ relations (e.g., richness--cluster mass, total luminosity--cluster mass), 
and to detect any possible cluster mass dependence on the redshift evolution of cluster galaxy properties (e.g., whether the BCG assembly history depends on cluster mass).
We emphasize that much better measurements of the  $M_{\rm gal}$--$M_{200}$ relation, including its slope and scatter, would provide strong constraints on the cluster formation models.
With  better information of cluster mass and density profile in hand, we will then be able to infer the contribution of pseudo-evolution \citep[][changes in cluster mass content simply due to the evolution in background or critical density of the Universe]{diemer13} to the way clusters evolve on the $M_{\rm gal}$--$M_{200}$ plane. 
Furthermore, we would be able to 
measure the stellar mass function (that is, number density per stellar mass interval), as well as radio luminosity function, which will facilitate comparisons between field and cluster studies.

Finally, in the current analysis we have only focused on one type of active galaxies (radio AGN).  With the data from the upcoming {\it eROSITA} mission, we will be able to apply the same technique to study the evolution of X-ray AGN in clusters.  It is also possible to include infrared-selected AGNs (e.g., using WISE data), and thus to have a comprehensive view of AGN population evolution in clusters.

%%%%%%%%%%%%%%%%%%%%%%%%%%%%%%%%%%%%%%%
%%%%%%%%%%%%%%%%%%%%%%%%%%%%%%%%%%%%%%%
\section{Summary}
\label{sec:summary}

In this paper we have presented a preview of what can be done with a uniformly selected cluster sample from the HSC-SSP survey, showing first-look results on the stellar mass assembly of BCGs at intermediate redshifts ($z=0.3-1.02$), tracing the evolution of the SMD, LD, and stellar mass surface density profiles for the red and blue populations, and, {\em for the first time}, studying the RLD and RAF in clusters out to $z\sim 1$.  
All these are carried out with the novel {\it top N} cluster selection, which is shown to allow us to faithfully follow the cluster galaxy evolution over cosmic time.
Another important aspect of our analysis is the {\it first} application of a machine-learning algorithm in estimating stellar mass of cluster galaxies,
which is shown to be unbiased and accurate, compared to the traditional template-fitting based methods.
Our main findings are summarized as follows.

1.~The typical mass of BCGs has increased by about 35\% from $z\approx 0.96$ to $z\approx 0.45$.  
This is about a factor of 2.2 lower than the prediction of a semi-analytic model \citep{guo13}, although the discrepancy is not significant given the scatter in  mass of our BCGs.

2.~The SMDs of clusters show noticeable evolution between $z=1$ and $z=0.3$.  
Between these two epochs, the abundance of galaxies (in terms of number per cluster) with stellar mass $M_{\rm star}\ge 10^{11}\,M_\odot$ doubles (similarly for $M_{\rm star}\ge 10^{10}\,M_\odot$ galaxies).
For low mass galaxies ($M_{\rm star}<10^{10}\,M_\odot$), the abundance also increases with time, with red galaxies showing more dramatic enhancement.
The stellar mass--cluster mass correlation is found to show no redshift evolution, which may point to substantial stellar mass loss during the hierarchical buildup of clusters.
Comparing to the field SMDs (after accounting for differences in densities), we find that the shape of the SMDs are similar for both red and blue populations, but clusters are over abundant in red galaxies. 

3.~Consistent with previous findings, the redshift evolution of the red galaxy population can be well described by a passively evolving stellar population forming at $z_f=3.5$.

4.~The stellar mass surface density profiles show a steady increase in amplitude with time, while keeping the shape roughly the same (and can be described by an NFW profile with low concentration $c_{200}=2.7-3.1$).  The mass density is dominated by the red galaxies.  

5.~Finally, we construct the RLD in clusters out to $z\sim 1$ for the first time, and 
find an over-abundance of radio galaxies in clusters compared to the field population.  In general, cluster galaxies at $z>0.77$ are about $1.5-2$ times more likely to be active in the radio (with 1.4 GHz radio power $\log P_{1.4}\ge 24.7$) compared to those in the lower-$z$ clusters.
The change in the relative abundance of massive red and blue galaxies ($M_{\rm star}\ge 10^{11}\,M_\odot$) could explain the nontrivial redshift evolution of the RAF.

We will continue to improve our analysis methods and refine our measurements.  With the much larger cluster sample to be delivered by the HSC-SSP survey in the near future, we will be able to obtain a comprehensive view of cluster galaxy evolution out to $z\sim 1.4$ or so.

%%%%%%%%%%%%%%%%%%%%%%%%%%%%%%%%%%%%%%%%
\section*{Acknowledgments}
%%%%%%%%%%%%%%%%%%%%%%%%%%%%%%%%%%%%%%%%

We thank an anonymous referee for comments that help improve the clarity of the paper.
YTL thanks Sesto Center for Astrophysics for the stimulating workshop ``The remarkable life of a BCG'' and Observat\'{o}rio Nacional in Rio de Janeiro for hospitality.  YTL is grateful to Andrey Kravtsov, Roderik Overzier, Robert Lupton, Yi-Kuan Chiang, and Greg Rudnick for helpful comments.
YTL acknowledges support from the Ministry of Science and Technology grants MOST 104-2112-M-001-047 and MOST 105-2112-M-001-028-MY3, and a Career Development Award  (2017-2021) from Academia Sinica.  
This project received financial support from the Conselho Nacional de Desenvolvimento Cient\'{i}fico e Tecnol\'{o}gico (CNPq) through grant 400738/2014-7.
YTL thanks IH and LY for constant support and encouragement.
This work was supported in part by World Premier International Research Center Initiative 
(WPI Initiative), MEXT, Japan, and JSPS KAKENHI Grant Number 26800093 and 15H05892.

The Hyper Suprime-Cam (HSC) collaboration includes the astronomical communities of Japan and Taiwan, and Princeton University. The HSC instrumentation and software were developed by the National Astronomical Observatory of Japan (NAOJ), the Kavli Institute for the Physics and Mathematics of the Universe (Kavli IPMU), the University of Tokyo, the High Energy Accelerator Research Organization (KEK), the Academia Sinica Institute for Astronomy and Astrophysics in Taiwan (ASIAA), and Princeton University. Funding was contributed by the FIRST program from Japanese Cabinet Office, the Ministry of Education, Culture, Sports, Science and Technology (MEXT), the Japan Society for the Promotion of Science (JSPS), Japan Science and Technology Agency (JST), the Toray Science Foundation, NAOJ, Kavli IPMU, KEK, ASIAA, and Princeton University. 

This paper makes use of software developed for the Large Synoptic Survey Telescope. We thank the LSST Project for making their code available as free software at  http://dm.lsst.org

The Pan-STARRS1 Surveys (PS1) have been made possible through contributions of the Institute for Astronomy, the University of Hawaii, the Pan-STARRS Project Office, the Max-Planck Society and its participating institutes, the Max Planck Institute for Astronomy, Heidelberg and the Max Planck Institute for Extraterrestrial Physics, Garching, The Johns Hopkins University, Durham University, the University of Edinburgh, Queen's University Belfast, the Harvard-Smithsonian Center for Astrophysics, the Las Cumbres Observatory Global Telescope Network Incorporated, the National Central University of Taiwan, the Space Telescope Science Institute, the National Aeronautics and Space Administration under Grant No. NNX08AR22G issued through the Planetary Science Division of the NASA Science Mission Directorate, the National Science Foundation under Grant No.~AST-1238877, the University of Maryland, and E\"{o}tv\"{o}s Lor\'{a}nd University (ELTE) and the Los Alamos National Laboratory.

This paper is based  on data collected at the Subaru Telescope and retrieved from the HSC data archive system, which is operated by Subaru Telescope and Astronomy Data Center, National Astronomical Observatory of Japan.

The Millennium Simulation databases used in this paper and the web application providing online access to them were constructed as part of the activities of the German Astrophysical Virtual Observatory (GAVO).

%%%%%%%%%%%%%%%%%%%%%%%%%%%%%%%%%%%%%%%%
%\bibliography{cosmology,refs}

\end{document}